\documentclass[3p,singlecolumn]{elsarticle}

\usepackage{tabularx,array}
\newcolumntype{Y}{>{\raggedright\arraybackslash}X}
\usepackage[dvipsnames]{xcolor}
\usepackage{subcaption} 
\usepackage{graphicx} 
\usepackage{booktabs}
\usepackage{circuitikz} 
\tikzset{>=latex} 
\usepackage{amsmath} 
\usepackage{amssymb}
\usepackage{cuted}
\usepackage{changepage}
\usepackage[linesnumbered,ruled]{algorithm2e}
\usepackage{graphicx}
\usepackage{url} 
\usepackage{circuitikz}

\usepackage{rotating}

\newcommand{\bmat}{\begin{bmatrix}}
\newcommand{\emat}{\end{bmatrix}}

\journal{Elsevier}

\begin{document}

\begin{frontmatter}


\title{Model predictive control lowers barriers to adoption of heat-pump water heaters: A field study}

\author[inst1,inst2]{Levi D. Reyes Premer}
\author[inst1,inst2,inst4]{Elias N. Pergantis}
\author[inst3]{Leo Semmelmann}
\author[inst1,inst2]{Davide Ziviani}
\author[inst1,inst2,inst5]{Kevin J. Kircher  \corref{correspondent}}

\affiliation[inst1]{organization={Ray W. Herrick Laboratories, School of Mechanical Engineering, Purdue University},
            city={West Lafayette},
            state={IN},
            postcode={47907}, 
            country={USA}}
\affiliation[inst2]{organization={Center for High Performance Buildings, Ray W. Herrick Laboratories, Purdue University},
            city={West Lafayette},
            state={IN},
            postcode={47907}, 
            country={USA}}
\affiliation[inst3]{organization={Karlsruher Institut für Technologie (KIT)},
            addressline={Kaiserstraße 12}, 
            city={Karlsruhe},
            postcode={76131}, 
            country={Germany}}
\affiliation[inst4]{organization={Trane Technologies, Residential R\&D Group},
            addressline={6200 Troup Hwy}, 
            city={Tyler},
            state={TX},
            postcode={75707}, 
            country={USA}}
\affiliation[inst5]{organization={Elmore Family School of Electrical and Computer Engineering, Purdue University},
            city={West Lafayette},
            postcode={47907}, 
            state={IN},
            country={USA}}

\cortext[correspondent]{Corresponding author: \texttt{kircher@purdue.edu}}

\begin{abstract}
Electric heat-pump water heaters (HPWHs) could reduce the energy costs, emissions, and power grid impacts associated with water heating, the second-largest energy use in United States housing. However, most HPWHs today require 240 V circuits to power the backup resistance heating elements they use to maintain comfort during large water draws. Installing a 240 V circuit can increase the up-front cost of a HPWH by half or more. This paper develops and field-tests the first control system that enables a 120 V HPWH to efficiently maintain comfort without resistance heating elements. The novel model predictive control (MPC) system enables pre-heating in anticipation of large water draws, which it forecasts using an ensemble of machine learning predictors. By shifting electrical load over time, MPC also reduces energy costs on average by 23\% and 28\% under time-of-use pricing and hourly pricing, respectively, relative to a 240 V HPWH with standard controls. Compared to the increasingly common practice in 120 V HPWHs of storing water at a constant, high temperature (60 $^\circ$C) to ensure comfort, MPC saves 37\% energy on average. In addition to demonstrating MPC's benefits in a real, occupied house, this paper discusses implementation challenges and costs. A simple payback analysis suggests that a 120 V HPWH, operated by the MPC system developed here, would be economically attractive in most installation scenarios. 
\end{abstract}


\end{frontmatter}

\section{Introduction}
\label{sec:intro}
\subsection{Heat-pump water heaters and their future}

\begin{figure*}
\centering
\includegraphics[width=\textwidth]{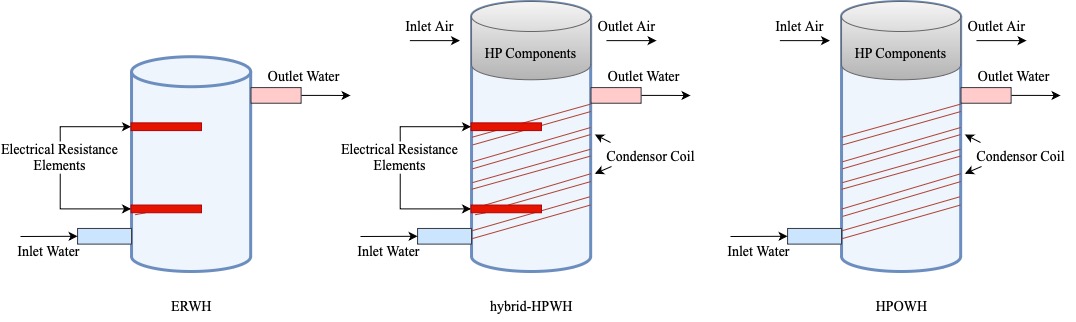}
\caption{Left: ERWH with only resistance heating elements. Center: Hybrid-HPWH with both a heat pump and heating elements. Right: HPOWH with a heat pump only. This paper focuses on HPOWH control.}
\label{fig:WH Types}
\end{figure*}

Heat pump technology is emerging as the future of domestic electric water heating. This trend is reinforced by federal and state policy \cite{CEC2024EnergyCode}, including a 2029 United States standard requiring most new electric storage water heaters to incorporate heat pump technology \cite{DOE2029HPWH}. These standards aim to improve the efficiency of water heating, the second-largest residential energy use \cite{RECS2020}. Today, electric resistance water heaters (ERWHs, illustrated at left in Figure \ref{fig:WH Types}) output at most one unit of heat transfer per unit of input electrical work. Heat-pump water heaters (HPWHs), by contrast, can output two to five units of heat transfer per unit of input electrical work. HPWHs do this by using electricity to transfer heat from the surrounding air into the water tank, rather than converting electricity to heat directly \cite{NRELHPWH2012, maguire_comparison_2013}. By improving energy efficiency, HPWHs can significantly reduce greenhouse gas emissions, energy costs, and impacts on power grids \cite{priyadarshan2024edgie, BILLERBECK2024117850}.

Despite the potential benefits of HPWHs, their adoption remains low, with about 3\% market share in the United States as of 2022 \cite{ENERGYSTAR2022_HPWHSales}. One significant barrier to HPWH adoption is high up-front costs \cite{ENERGYSTAR2022_HPWHSales, ENERGYSTAR2024_HPWH}. Today, many HPWH models are hybrids that have both a heat pump and electric resistance heating elements (the middle configuration in Figure \ref{fig:WH Types}). To run the heating elements, hybrid-HPWHs typically require a 240 V outlet, wiring rated at 30 A, and a free slot in the circuit breaker panel. In homes that lack some or all of this electrical infrastructure, installing a hybrid-HPWH requires electrical work that can cost hundreds to several thousand dollars \cite{REWIREAMRERICA_HPWH, SATREMELOY2024110939}. In the 21\% of United States housing that has electrical panels and service rated at 100 A or below, installing a hybrid-HPWH may require panel or service upgrades that cost an additional \$2,000 to \$10,000 \cite{pergantis_current, pergantis_current_APEN}. Some households may struggle to justify these electrical upgrades, which can cost several times more than the hybrid-HPWH itself. 

\textcolor{black}{These barriers to adoption are unique to regions where domestic water heating uses different equipment from space conditioning, and where two outlet standards exist -- a lower voltage (e.g., 120 V) and a higher voltage (e.g., 240 V). Many European regions do not face these challenges, as domestic water heating is often integrated with space conditioning and there is a single standard outlet voltage (220--240 V). However, the strategies developed in this paper may still offer value in European contexts, as they enable load-shifting and improved thermal comfort.} 

Recently, electrical infrastructure challenges in the United States have motivated the manufacture of heat-pump-only water heaters (HPOWHs, shown at right in Figure \ref{fig:WH Types}), which do not have resistance heating elements. HPOWHs plug into standard 120 V outlets and circuits rated at 15 A, reducing or eliminating the need for electrical upgrades when transitioning from fossil-fueled water heating. However, today's HPOWHs \cite{osti_885625, efficiencyfirstca_heat_pump_water_heaters} may have difficulty maintaining comfortable water temperatures, as they take longer than hybrid-HPWHs to reheat the tank once the stored hot water is depleted. 

This paper is the first to develop and field-test control algorithms that enable HPOWHs to maintain comfortable water temperatures by proactively heating in anticipation of large water draws. The algorithms developed here could lower barriers to HPWH adoption by eliminating the need for electrical upgrades while alleviating concerns about thermal comfort. Several studies have used MPC on HPWHs to reduce peak demand, greenhouse gas emissions, and energy costs, and to enhance demand flexibility \cite{mande2022timing, shen2021data, dela2021supervisory, BAUMANN2023112923, Jin2014}. These studies demonstrate the advantages of hybrid-HPWHs over fossil-fueled water heaters and ERWHs. Transitioning to HPOWHs has the potential to increase these advantages while also reducing installation costs, a barrier for hybrid-HPWHs. However, the studies cited above do not address the thermal comfort challenges remaining with HPOWHs.

HPOWH control is an emerging research area. As a foundational contribution to this research area, Baumann et al. \cite{BAUMANN2023112923} conducted a comprehensive laboratory study to evaluate the performance of MPC relative to traditional hysteresis control for HPOWHs under simulated water draw conditions. Their study demonstrated the potential for energy and cost savings with MPC. Although comfort was not a primary focus of the paper, their results highlighted occasional comfort issues. Their implementation involved a considerable number of sensors and relied on a data-heavy water draw forecast. This approach is likely not practical in field implementations, as homes typically have limited data availability and additional sensors can significantly increase installation costs. This paper builds on the work of Baumann et al. by transitioning from controlled laboratory experiments to field demonstrations in a real, occupied house, capturing unique disturbances absent in laboratory settings. This paper also develops a sensing and communication architecture that can be deployed at comparatively low cost and enables data-driven prediction and control algorithms that are robust to the limited data typically encountered in the field.

\subsection{Contributions of this paper}

This paper makes four main contributions to the research literature. First, this paper uses field data from an occupied house to demonstrate thermal comfort issues associated with HPOWHs in typical households. The analysis focuses on the challenges associated with maintaining thermal comfort without using resistance heating elements.

Second, this paper develops a convex MPC formulation to enhance comfort in HPOWHs, emphasizing computational efficiency to allow implementation on low-cost microprocessors. The formulation includes a second-order linear model that captures water temperature dynamics more accurately than the first-order linear models often used today. This paper develops parameter estimation procedures for the second-order model. The model fits empirical data well and integrates readily with practical control algorithms that adjust temperature set-points via manufacturers' application programming interfaces (APIs). 

Third, this paper develops a novel ensemble forecasting algorithm to predict domestic hot water draws with limited training data. The algorithm combines two machine learning models and resampled historical water draw data to generate rolling 24-hour forecasts, capturing usage patterns and enhancing the predictive capability of MPC.

Fourth, this paper demonstrates the full system described above -- including the sensing, communication, and actuation infrastructure, the ensemble water draw prediction algorithm, and MPC -- in an occupied home. As far as the authors know, this is only the second field demonstration of any advanced control strategy for any type of HPWH and the first for HPOWHs. The field demonstration results in this paper could build confidence that advanced HPWH control can deliver attractive benefits at low deployment costs, moving the technology closer to real-world adoption. By discussing solutions to a number of practical Internet-of-Things (IoT) challenges, this paper could also enable future researchers to more easily demonstrate their own HPWH control algorithms in the field.

\subsection{Outline of this paper}

This paper is organized as follows. Section \ref{sec:related work} reviews related work from industry and academia. Section \ref{sec:methodology} describes the methods used to develop the MPC algorithm: Modeling the water heater, forecasting water draws, and formulating the optimization problem. Section \ref{sec:Test House} describes the test house, including the IoT infrastructure, the water heater, the occupants, and comfort issues with HPOWHs. Section \ref{sec:Test House} also presents the accuracy of the thermal model and water draw forecasts as implemented in the test house. Section \ref{sec:results} presents control results from the field test. Section \ref{sec:discussion} discusses implementation costs, energy savings, and a simple payback analysis. Section \ref{sec:discussion} also discusses limitations of this paper and directions for future work. Section \ref{sec:conclusion} concludes the paper.

\section{Related work}
\label{sec:related work}

This section reviews past work on residential water heater control. First, it outlines the rationale for investigating residential water heaters as tools for load-shifting. Next, it examines a commercially available demand response controller, evaluating its potential and limitations. The progression of water draw forecasting methods is then explored, highlighting advanced techniques but noting their lack of practicality for widespread use. The review transitions to ERWH studies, which laid the groundwork for this field, before shifting focus to HPWH studies, primarily conducted in simulation. Finally, experimental HPWH studies are discussed in detail, highlighting their contributions and pinpointing the gaps this present work seeks to address. 

\subsection{Grid-interactive water heating}
\label{sec:ERWHandHPWH}

Electric water heating, including ERWHs and HPWHs, presents unique challenges and opportunities \cite{Bastian2022}. Some challenges include additional strain on the grid \cite{TARROJA2018522,priyadarshan2024edgie}, costly electrical upgrades \cite{REWIREAMRERICA_HPWH}, and potential reliance on carbon-intensive electricity in regions with significant fossil-fueled generation \cite{Decarb2050DOE}. As water heaters with tanks inherently have thermal storage, there are various opportunities to overcome these challenges. Thermal storage enables load-shifting, which can improve grid stability \cite{ZHANG2019709, LACROIX19991313}, lower carbon emissions \cite{DELAROSA202583}, reduce operational costs \cite{Jin2014}, and enable better utilization of renewable energy sources \cite{EARLE2023120256, ZHAO2024119026,YANG2021114710}. The potential impact of grid-interactive electric water heaters in the United States is reviewed by Silvestre et al. \cite{DISILVESTRE2023113425}.

\subsection{Industry's demand response controls}
\label{sec:IndustrySolutions}

Water heater demand response programs could save utilities millions of dollars each year \cite{cta2045report2018}. In today's water heater demand response programs, load-shifting typically involves raising water heater set-points during off-peak hours (referred to as `load-up') and lowering them during peak hours (`shed') \cite{CTA2045B2022}. Commercial products like CTA-2045 determine load-up or shed times based on utility schedules or direct signals, without automatically considering occupant water draw trends \cite{CTA2045B2022}.

In a study of 191 customers using CTA-2045 on ERWHs and hybrid-HPWHs, around 40\% reported cold water draws, leading two participants to withdraw due to comfort issues \cite{cta2045report2018}. Despite this, 80\% of participants expressed satisfaction, primarily due to monetary incentives and perceived contributions to enabling a cleaner and more robust grid. For customers less motivated by environmental concerns, delivering both comfort satisfaction and monetary incentives could likely reduce drop-out rates in demand response programs. Predictive controls could improve comfort during demand response events and enable more sophisticated demand response signals, increasing the value of water heater demand response to power grids \cite{DISILVESTRE2023113425}.

\subsection{Forecasting domestic hot water draws}
\label{sec:forecasting}

Predictive water heater controls require forecasting of water draws, a research area that has significantly advanced in recent years. Early work by Jordan et al. \cite{Jordan2001} introduced statistical methods to generate generalized domestic hot water profiles based on household characteristics and time-of-use information. These approaches, however, lacked granularity to capture individual household behaviors. Subsequent studies collected extensive data from multiple households to better understand daily domestic hot water demand, yet they required heavy data inputs and didn't fully customize predictions for individual homes \cite{RITCHIE2021110727, EDWARDS201543, shen2021data}. Ritchie et al. \cite{RITCHIE2021110727} also demonstrated that hot water demand can vary substantially between homes, depending on factors like household size, region, season, and lifestyle. All of these models required heavy amounts of data and their ability to customize the prediction for a single household's behavior was not thoroughly evaluated. 

Cao et al. \cite{Cao2019} learned water draw behavior using a support vector machine. The support vector machine fit individual resident behavior but relied on occupant participation to provide data on when they took showers. Work from Pflugradt et al. introduced a behavioral model that simulated water and energy use based on the psychology of users \cite{PFLUGRADT2017655}. The model was effective but required potentially intrusive data inputs, making it less practical for larger-scale applications. Gelazanskas and Gamage \cite{Gelazanskas2015} explored time series models, specifically seasonal autoregressive integrated moving average and decomposition of time series by Loess model, to forecast residential hot water demand for individual households. They found that incorporating seasonal decomposition improved prediction accuracy for individual households, with the best model performance from a combination of the autoregressive and Loess models. For each household, the method required one full year of historical data, making practical implementation difficult. 

The development of a water draw forecast model for an individual household with limited data has received little attention in the research literature. Improvements in the practicality of these forecasts will help close the gap between research and industry applications for demand response water heaters.

\subsection{ERWH studies}
\label{sec: ERWHstudies}

Research has focused on applying predictive control algorithms to ERWHs to improve grid interactivity and facilitate demand response \cite{DISILVESTRE2023113425, CLIFT2023126577}. Many studies have developed model-based control methods to achieve peak curtailment to preserve the grid \cite{DISILVESTRE2023113425, RITCHIE2024123421, Buechler_Goldin_EWH1_2024}. Some studies developed low-level controllers that directly manipulated the resistance heating elements \cite{Buechler_Goldin_EWH1_2024}, while others developed high-level controllers that adjusted water temperature set-points (the approach taken in this paper) \cite{shen2021data, RITCHIE2024123421, 2nodeHmodel_DIAO1012, CLIFT2023126577}. Data-driven modeling methods have been explored to enable demand response \cite{LUO2024110724, Lin_Li_Xiao_2017}. These methods strive to increase the scalability of advanced controllers for water heaters. Using predictive control, the integration of ERWHs with renewable energy, such as photovoltaic systems, has also been studied to enable cost-effective energy storage and minimize wasted photovoltaic generation \cite{ELBAKALI2024118190, GAONWE2022}.

\subsection{Simulation-based HPWH studies}
\label{sec:simulated HPWH studies}

Studies on HPWHs, particularly in grid interactivity and demand response, have been more limited compared to ERWHs. Most grid-interactive HPWH research has been limited to simulations. Jin et al. \cite{Jin2014} conducted one of the first extensive studies, developing a mixed-integer MPC system specifically for hybrid-HPWHs. Their simulations demonstrated notable energy and cost savings, as well as enhanced thermal comfort, highlighting the benefits of predictive control approaches.

Additional research focuses on reducing operating costs and carbon emissions through predictive controls such as MPC \cite{mande2022timing, dela2021supervisory, DELAROSA202583, DELAROSA2025_Rule} and using HPWHs for secondary frequency control \cite{en12030411}. These studies often employ time-of-use rate structures and consider greenhouse gas emissions from electricity generation. Amasyali et al. \cite{amasyali_deep_2021} introduced a reinforcement learning agent that doesn't rely on future water draws, resulting in behavior similar to rule-based control methods like the CTA-2045 \cite{CTA2045B2022}. Their simulations suggest that reinforcement learning with predictions may offer greater savings than MPC optimized with genetic algorithms, assuming perfect water draw forecasts and extensive sensing capabilities.

Simulation-based research on HPWH control often assumes perfect water draw forecasts and extensive sensing capabilities, which may not be available in practice. Our research addresses these limitations by focusing on practical implementation with limited data and minimal sensing equipment.

\subsection{Experimental HPWH studies}
\label{sec:experimentalHPWH}

Only three experimental studies -- one in a laboratory and two in the field -- have been conducted to date, to the authors' knowledge. By providing empirical data and real-world insights, experimental results validate and enhance the credibility of model predictions. Experimental studies not only strengthen confidence in the research findings but also facilitate the translation of theoretical advancements into industry practice.

Baumann et al. \cite{BAUMANN2023112923} made a significant contribution with their extensive laboratory experiment comparing MPC to traditional hysteresis control for HPWHs, using simulated water draw profiles from Ritchie et al. \cite{RITCHIE2021110727}. Baumann et al. experimented on a HPWH under Austria's day-ahead energy market prices. Their MPC algorithm used a probability-based water draw prediction derived also from \cite{RITCHIE2021110727}, relying on four months of data collected from 77 homes. The results demonstrated substantial energy and cost savings -- 11.38\% and 23.96\%, respectively -- compared to maintaining a constant high set-point. However, their findings also indicated some discomfort during controller operation. While this method provides valuable insights into the potential of MPC-operated HPWHs, the evaluation lacked real-world water draw disturbances typical in residential settings, and the forecasting approach may be impractical for field implementation. Additionally, the controller required a considerable number of sensors. Nevertheless, their innovative approach offers valuable insights and lays a foundation for future research into more scalable and practical control strategies.

Yin et al. \cite{yin_data-driven_2024} experimentally evaluated data-driven predictive control of a domestic water heater, with heat supplied by a central air-to-water heat pump. The control technique, based on behavioral systems theory with signal-matrix modifications, directly uses past input-output data to predict the response of water temperatures and power draws to control actions. This data-driven representation of the system enables fast deployment of predictive control, as it does not require developing thermal models. While providing energy savings, this control framework faced limitations in fully maintaining comfort due to uncertainties in water draw forecasts.
Starke et al. managed a community's hybrid-HPWHs with mixed-integer MPC to reduce peak electricity demand \cite{Starke2020}. Field tests were conducted in a `Smart Neighborhood' \cite{StarkeNeighbor2019}, consisting of 62 homes. Starke et al. designed an IoT infrastructure for the entire community using a multi-agent system network \cite{Starke2020}. A cloud-based virtual machine solved an MPC optimization problem for each home's water heater. Each home was equipped with a flow meter to measure hot water draws. A time-of-use rate structure was provided by the utility. In their analysis of one day of operation for two homes, Starke et al. showed that the controller pre-heated water before peak pricing periods and shifted load during times of small water draws. However, a thorough analysis of the water heaters' energy savings and occupant comfort was not presented. Nonetheless, their work provides important insights into the IoT infrastructure required to coordinate multiple HPWHs in a community with centralized control.

Building on the three studies mentioned above, this paper offers a comprehensive analysis of a comfort-driven predictive control solution for HPOWHs, focusing on the unique challenges faced in real-world field settings. Unlike previous studies, our research demonstrates a practical implementation of predictive control with limited data -- just 30 days of water draws -- and minimal sensing equipment, making it feasible for real-world applications. The proposed system uses an ensemble forecasting algorithm that combines multiple machine-learning models. Additionally, a second-order semi-physical model of the water temperature dynamics was created to enable the controller to run efficiently on low-cost devices such as a microprocessors or via cloud-based platforms. This research aims to bridge the gap between simulation and practical application, demonstrating the potential for cost savings, comfort assurance, and enhanced grid flexibility with HPOWHs.

\section{Methods}
\label{sec:methodology}

This section outlines scalable methods for improving residential water heater control. Sections \ref{sec:Modeling} and \ref{sec: ParameterTuning} develop a two-node thermal model that captures tank stratification and enables tractable set-point control of the water heater. Section \ref{sec:ForecastModel} develops an ensemble of methods for forecasting hot water draws. Section \ref{sec:ControllerDesign} incorporates the tank model and water draw forecasts into a predictive control framework to minimize operating costs under various rate structures while ensuring comfortable outlet water temperatures.

\subsection{Modeling}
\label{sec:Modeling}

There are various ways to model the temperature distribution within a water tank. For advanced control applications, researchers often use single-node models that assume the water temperature is spatially uniform throughout the tank \cite{Starke2020, mande2022timing, BAUMANN2023112923, shen2021data}. While single-node models are simple and straightforward to implement, they fail to account for the stratification of water temperatures within the tank. Stratification is particularly pronounced during water draws, as cooler water enters at the bottom of the tank while hot water is drawn from the outlet at the top. Manufacturers typically design water heaters to promote stratification, as drawing water from the top of a stratified tank helps ensure consistently warm and comfortable outlet temperatures. Consequently, single-node models fail to accurately represent outlet water temperatures in practice \cite{Nash_Badithela_Jain_2017, Xu2024, Jin2014}. This limitation is particularly critical in HPOWHs, where the heat transfer rate is constrained, making the water temperature at the top of the tank a key factor in ensuring comfortable outlet temperatures. Multi-node models overcome this limitation of single-node models by capturing stratification effects.

The number of nodes in a multi-node water tank model is a design choice. While increasing the number of nodes may improve model fidelity, using a multi-node model in practice requires measuring or estimating each node temperature \cite{Nash_Badithela_Jain_2017}. Since residential water heaters typically have temperature sensors near the top and bottom of the tank, a two-node model offers a practical representation aligned with available measurements \cite{Buechler_EWH2_2024}. Higher-order models can be implemented using state estimators, although this adds complexity to the control system \textcolor{black}{and requires tuning additional parameters, potentially undermining scalability}. Compared to a single-node model, a two-node model better represents tank stratification and more accurately \textcolor{black}{predicts the outlet water temperature \cite{Jin2014}. This was confirmed during our initial model development. The difference in outlet water temperature predictions between the single- and two-node models is most evident during water draws, when the temperature near the bottom of the tank decreases due to cold water inflow. This effect causes the single-node model to predict a corresponding drop in the lumped tank temperature, which is its predictor of the outlet water temperature. In contrast, the two-node model allows the lower-node temperature to decrease without significantly affecting the upper-node temperature (the two-node model's predictor of the outlet water temperature), reflecting the stratification behavior of the physical system.} The two-node model developed here draws inspiration from Jain et al. \cite{Nash_Badithela_Jain_2017} and Jin et al. \cite{Jin2014}. 

Two approaches exist for two-node modeling of water tanks. Both approaches model two stacked columns of water (referred to here as `nodes') separated by a thin thermocline. Both approaches also assume that the water temperature is spatially uniform within each node. The first modeling approach assumes that the lower and upper-node temperatures are time-invariant, but treats the position of the thermocline as dynamic \cite{2nodeHmodel_DIAO1012}. The second modeling approach assumes that the thermocline position is time-invariant, but treats the temperatures of the lower and upper nodes as dynamic \cite{Jin2014}. Given that the water heater control approach developed here centers on set-point temperature, the second modeling approach -- fixing the thermocline height -- was adopted for this study. This approach enables the optimization problem to directly output a set-point temperature that can be communicated to the water heater without additional processing.

\begin{figure}
\centering
\begin{circuitikz}[scale=0.8]
    \ctikzset{bipoles/length=0.75cm}
    
    \draw[thick] (-2,-1.5) -- (-2,3.5);
    \draw[thick] (2,-1.5) -- (2,3.5);
    \draw[|-|] (-4,-1.5) -- (-4,3.5);
    \node[right] at (-4,1) {$h$};
    \draw[|-|] (-3.5,-1.5) -- (-3.5,0.5);
    \node[right] at (-3.5,-0.5) {$h_\text{thrm}$};
    \draw[gray, dashed] (-4,3.5) -- (-2,3.5);
    \draw[gray, dashed] (-4,-1.5) -- (-2,-1.5);
    \draw[gray, dashed] (-3.5,0.5) -- (-2,0.5);
    
    \draw[thick] (-2,3.5) arc (180:360:2 and 0.5);
    \draw[thick] (-2,3.5) arc (180:360:2 and -0.5);
    \node at (0,3.5) {$z = 1 - h_\text{thrm} / h$};
    
    \draw[thick] (-2,-1.5) arc (180:360:2 and 0.5);
    
    \draw [dashed,gray] (0,0.5) ellipse (2 and 0.5);
    
    \node[above right] at (0,1.5) {$T_u$};
    \draw (0,1.5) to[R,n=Rh,*-] (4,1.5) to (4,0.5);
    \node[below right] at (Rh.s) {$R_a/z$};
    
    \draw (0,-0.5) to[R,n=Rc,*-] (4,-0.5) to (4,0.5);
    \node[below right] at (2,-0.5) {$R_a/(1-z)$};
    \node[below right] at (0,-0.5) {$T_\ell$};
    
    \draw (0,1.5) to[R,n=Rsl,-] (0,-0.5);
    \node[right] at (Rsl.s) {$\ R_{u\ell}$};
    
    \draw (4,0.5) to[battery1,*-] (5.0,0.5) node[ground] {};
    \node[left] at (4,0.5) {$R_a$};
    
    \draw ++(-1.5,1.5) node[ground] {} to[C] (0,1.5);
    \node[above] at (-0.75,1.75) {$z C$};
    
    \draw ++(-1.5,-0.5) node[ground] {} to[C] (0,-0.5);
    \node[below] at (-0.75,-0.75) {$(1-z)C$};
        
    \draw[->] (2.5,2.5) -- (1.5,2.5);
    \node[right] at (2.5,2.5) {$\lambda q$};
    \draw[->] (2.5,-1.5) -- (1.5,-1.5);
    \node[right] at (2.5,-1.5) {$(1-\lambda) q$};


\end{circuitikz}
\caption{In a two-node water tank model, a thermocline separates the upper and lower water columns. Within each column, temperatures are spatially uniform but time-varying.}
\label{fig:2nodemodel}
\end{figure}
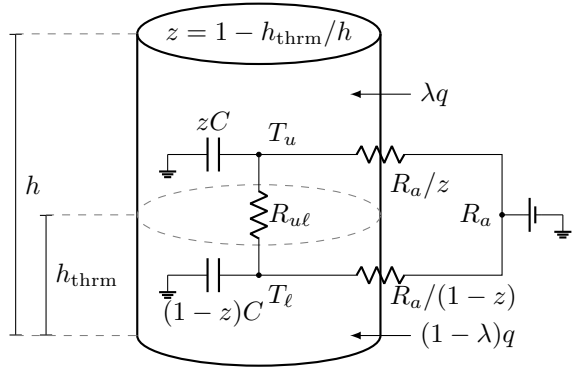

Figure \ref{fig:2nodemodel} illustrates the two-node water tank model used here. The water is treated as an incompressible fluid, and a thermal circuit representation is used to define the governing differential equations \eqref{T_u Energy Balance} and \eqref{T_L Energy Balance}. The thermal circuit is analogous to an electrical circuit, with temperature playing the role of voltage, heat playing the role of charge, and thermal capacitances and resistances playing the roles of electrical capacitors and resistors. 

The tank is divided into two internally well-mixed nodes, with the dimensionless parameter $z = 1 - h_\text{thrm} / h$ defining the fraction of the total tank height $h$ (m) corresponding to the upper node. Here $h_\text{thrm}$ is the height of the thermocline that separates the upper and lower nodes. The total thermal capacitance $C$ (kWh/kg) of the water, including the upper and lower nodes, is the product of the water's density, specific heat capacity, and volume. The individual thermal capacitances of the upper and lower nodes are $zC$ and $(1-z)C$, respectively. The rate of heat transfer from the heat pump, $q$ (kW), is split between the two nodes via the dimensionless factor $\lambda$, such that the upper node receives $\lambda q$ and the lower node $(1-\lambda)q$.

Heat transfer between the upper and lower nodes is governed by the thermal resistance $R_{u\ell}$ ( $^\circ$C/kW), which models conductive heat transfer within the water as well as mass transfer due to thermal buoyancy effects. The upper- and lower-node temperatures are $T_u$ and $T_\ell$ ($^\circ$C). The ambient environment is modeled as a thermal reservoir with infinite capacitance, analogous to the ``ground'' in an electrical circuit. The heat transfer between the water in the tank and the ambient air is governed by the thermal resistance of the entire tank wall, $R_a$ (kW/ $^\circ$C). The resistances between each node and the ambient are scaled according to each node’s relative height within the tank. 

With these elements in place, applying Kirchhoff's laws to the thermal circuit yields the differential equations that govern the water temperature dynamics:
\begin{equation}
\begin{aligned}
&zC\dot{T}_u(t) = \dfrac{T_\ell(t) - T_u(t)}{R_{u\ell}} + \dfrac{z(T_a(t) - T_u(t))}{R_a}  \\
&\quad + \lambda q(t) + \dot m(t) c_p (T_\ell(t) - T_u(t)) 
\end{aligned}
\label{T_u Energy Balance}
\end{equation}
for the upper node and 
\begin{equation}
\begin{aligned}
&(1 - z)C\dot{T}_\ell(t) = \dfrac{T_u(t) - T_\ell(t)}{R_{u\ell}} \\
&\quad + \dfrac{(1 - z)(T_a(t) - T_\ell(t))}{R_a} + (1 - \lambda) q(t) \\
&\quad + \dot m(t) c_p (T_c(t) - T_\ell(t))
\end{aligned}
\label{T_L Energy Balance}
\end{equation}   
for the lower node. The left-hand sides of Equations \eqref{T_u Energy Balance} and \eqref{T_L Energy Balance} represent the rates of change of internal energy for the upper and lower nodes. The right-hand sides include terms that model (from left to right) heat and mass transfer between the two nodes, heat transfer between each node and the ambient air, heat transfer from the heat pump's condenser, and mass transfer associated with hot water withdrawals from the upper node and cold water intake to the lower node. Under the incompressible substance model, heat transfer associated with mass transfer is the product of the mass flow rate $\dot m$ (kg/h), the specific heat capacity $c_p$ (kWh/(kg$\cdot^\circ$C)) of water, and the relevant temperature difference. The mass flow rate is assumed to be the same for the water entering and exiting each node. For the lower node, the relevant temperature difference is between the cold water inlet temperature $T_c$ ($^\circ$C) and the lower-node temperature $T_\ell$. For the upper node, the relevant temperature difference is between the lower-node temperature $T_\ell$ and the upper-node temperature $T_u$. Buoyancy effects are not explicitly included in the formulation in order to preserve linearity. However, the thermal resistance $R_{u\ell}$ between the upper and lower nodes implicitly captures some of these effects. Moreover, it is rare in practice for the upper layer’s temperature to fall below that of the lower layer, a condition under which buoyancy-driven mixing would be most significant. 

The heat pump's rate of heat transfer output $q$ and electric power input $P$ (kW) are related through the coefficient of performance (COP) $\eta$: $q(t) = \eta P(t)$. The COP, $\eta$, is a function of ambient conditions and condensing temperature. \textcolor{black}{The COP also depends on equipment sizing and the condenser’s interaction with water. As a result, different HPWHs can have different COP values under the same ambient conditions and condensing temperatures. For a given HPWH, however, these design choices are fixed and define its COP curve.} Manufacturers typically provide a simplified COP curve based solely on the condensing temperature, which we obtained access to. A linearization of this relationship using a Taylor series expansion was explored; however, this introduced an offset at zero, implying ``free heat'' when no heating was required, thus misleading the MPC optimization. To mitigate this misleading zero offset, we assume that the COP is constant, which is reasonable if the temperatures of the water and of the air surrounding the tank are approximately constant. \textcolor{black}{In practice, the condensing temperature does vary, and the constant COP assumption is not accurate under all conditions. Tuning the constant-COP value is therefore important to minimize prediction errors. A tuning method is introduced in Section \ref{sec: ParameterTuning}.} While the constant-COP assumption limits the controller's explicit awareness of reduced COP at higher condensing temperatures, the optimization formulation in Section \ref{sec:ControllerDesign} implicitly discourages high condensing temperatures through other cost terms, such as increased power consumption and thermal losses. MPC's feedback mechanism also compensates for discrepancies between the constant-COP model and the actual physics. \textcolor{black}{For example, if the constant COP is higher than the realized COP at a given time, the heat input will be less than modeled. The feedback mechanism in MPC will then adjust accordingly to this modeling error at the next time step.}

The governing equations \eqref{T_u Energy Balance} and \eqref{T_L Energy Balance} can be written as
\begin{equation}
\begin{bmatrix}
    \dot{T}_u(t) \\ 
    \dot{T}_\ell(t) \\
\end{bmatrix} = \tilde A(t) \begin{bmatrix}
    T_u(t) \\ 
    T_\ell(t) \\
\end{bmatrix} + \tilde B q(t) + \tilde w(t) . 
\label{eq:state_space}
\end{equation}
Here $\tilde A \in \mathbf R^{2 \times 2}$ and $\tilde B \in \mathbf R^2$ are continuous-time system matrices, $T_u$ and $T_\ell$ are state variables, $q$ is the control input or action, and $\tilde w \in \mathbf R^2$ is an additive disturbance. The system matrices and disturbance are defined by
\begin{equation}
\begin{aligned}
\tilde A_{11}(t) &= - \frac{1}{z C} \left( \frac{1}{R_{u\ell}} + \frac{z}{R_a} + \dot m(t) c_p \right) \\
\tilde A_{12}(t) &= \frac{1}{z C} \left( \frac{1}{R_{u\ell}} + \dot m(t) c_p \right) \\
\tilde A_{21} &= \frac{1}{(1-z) C R_{u\ell} } \\
\tilde A_{22}(t) &= -\frac{1}{(1-z) C} \left(\dot m(t) c_p + \frac{1}{R_{u\ell}} + \frac{1-z}{R_a} \right) \\
\end{aligned}
\end{equation}
\begin{equation}
\tilde B = \frac{1}{C} \begin{bmatrix}
\lambda / z \\
(1-\lambda) / (1-z)
\end{bmatrix} 
\end{equation}
\begin{equation}
\tilde w(t) = \frac{1}{C} \begin{bmatrix}
T_a(t) / R_a \\
T_a(t) / R_a + \dot m(t) c_p T_c(t) / (1-z) 
\end{bmatrix} .
\end{equation}
Assuming $\dot m$, $q$, $T_a$, and $T_c$ are piecewise constant over each time step, the continuous-time system \eqref{eq:state_space} corresponds exactly to the discrete-time system
\begin{equation}
\begin{bmatrix} T_u(k+1) \\ T_\ell(k+1) \end{bmatrix} = A(k)\begin{bmatrix} T_u(k) \\ T_\ell(k) \end{bmatrix} + B(k) q(k) + w(k) . \label{eq:state} 
\end{equation}
Here $k$ indexes time steps, the notation $T_u(k)$ indicates $T_u(t_k)$, and we use similar temporal notation for the system matrices and input signals. Straightforward calculations show that $\tilde A$ is invertible. Therefore, the discrete-time system matrices and disturbance are defined by
\begin{equation}
\begin{aligned}
A(k) &= e^{\tilde A(t_k) \Delta t} \\
B(k) &= \tilde A(t_k)^{-1} (e^{\tilde A(t_k) \Delta t} - I) \tilde B \\
w(k) &= \tilde A(t_k)^{-1}(e^{\tilde A(t_k) \Delta t} - I)\tilde w(t_k) ,
\end{aligned}
\end{equation}
where $\Delta t$ (h) is the time step and $I$ is the $2 \times 2$ identity matrix.

\subsection{Parameter tuning}
\label{sec: ParameterTuning}

The fundamental model parameters are $R_{u\ell}$, $R_a$, $C$, $\eta$, $\lambda$, and $z$. The parameters $R_a$ and $C$ are straightforward to define given the tank geometry, the density and specific heat of water, and insulation information from manufacture specification sheets. This section focuses on the remaining parameters $R_{u\ell}$, $\eta$, $\lambda$, and $z$, which are more challenging to define.

Determining the nondimensionalized thermocline height $z \in [0,1]$ and the fraction $\lambda \in [0,1]$ of the heat pump's rate of heat transfer output that enters the upper node is challenging, as the model in Section \ref{sec:Modeling} is a time-invariant representation of the true system's time-varying dynamics. For example, the dynamics during water draws differ substantially from the dynamics when the tank is fully charged and water temperatures drift slowly due to mixing and to conduction through the tank wall. It is also challenging to choose a constant value of the COP $\eta > 0$ that reasonably approximates the true COP, which varies with the temperatures of the upper node, lower node, and ambient air. The resistance between the two nodes,
\begin{equation}
R_{u\ell} = \frac{h_s}{k_w A_{\text{WH}}} ,
\label{eq: strat_layer}
\end{equation}
depends on the (unknown) thickness $h_s$ (m) of the stratification layer, as well as the tank's (known) horizontal cross-sectional area $A_{\text{WH}}$ ($m^2$), and the (known) thermal conductivity of water, $k_w = 0.63$ W/m/ $^\circ$C. The challenge in defining $R_{u\ell}$ is to find a reasonable value of $h_s$.

To address these challenges, parameters can be tuned by comparing the model predictions to temperature and power measurements from the real system. As the number of tunable parameters is small, we use a joint grid search over $\eta \in \{1, 1.1, \dots, 5\}$, $h_s \in \{0.005, 0.015, \dots, 0.1\}$, and $\lambda$ and $z \in \{0,0.1,\dots,1\}$. The grid search aims to minimize the unweighted sum of the mean relative absolute errors in predicting the power $P$ and the upper tank temperature $T_u$. The tuned parameters are sensitive to the measurement data used in the grid search. As discussed in Section \ref{sec:Test House}, using at least two weeks of data with typical hot water draw profiles captured sufficient variations in water heater dynamics. Seasonal changes, such as lower inlet water temperatures in winter compared to summer, may also influence parameter values. In climates with significant seasonal variation, re-tuning parameters each winter and summer may be warranted.

\subsection{Water draw forecasting}
\label{sec:ForecastModel}

The reliability and performance of MPC algorithms depend on the quality of the underlying forecasting model \cite{Bartolucci2019}. However, most studies focused on MPC algorithms for water heaters make the unrealistic assumption of perfect forecasts \cite{amasyali_deep_2021, LydenTuohy2022} or rely forecasting approaches that have significant room for improvement \cite{shen2021data}. Hence, this study aims to contribute to the understanding of short-term hot water demand forecasting by elaborating on a state-of-the-art forecasting pipeline. The analysis focuses on forecasting performance across various horizons to construct an ensemble model. The methodology encompasses feature engineering, model selection, and evaluation metrics. The final model forecasts the water draws over the next 24 hours at five-minute resolution, corresponding to a forecasting horizon of $J = 288$.

\subsubsection{Feature engineering} 

The given measurements for the forecasting model include measurements of the past hot water flow, the upper tank temperature, and the lower tank temperature. Initial experiments suggested that the forecast quality can be improved by incorporating certain additional features. First, a sine and cosine transformation is added for the hour of each time step $k$ in Equations \ref{eq:sinhour} and \ref{eq:coshour}:
\begin{equation}
\sin_h(k) = \sin( 2 \pi \cdot \text{hour}(k) / 24 )
\label{eq:sinhour}
\end{equation}
\begin{equation}
\cos_h(k) = \cos( 2 \pi \cdot \text{hour}(k)/ 24 ) .
\label{eq:coshour}
\end{equation}
Previous studies have shown that this transformation potentially improves the forecast since it yields a cyclical representation of time \cite{semmelmann2022load}. Similar transformations are applied to the day of the week in Equations \ref{eq:sinweekday} and \ref{eq:cosweekday} for each $k$:
\begin{equation}
\sin_w(k) = \sin(2 \pi \cdot \text{weekday}(k) / 7 )
\label{eq:sinweekday}
\end{equation}
\begin{equation}
\cos_w(k) = \cos( 2 \pi \cdot \text{weekday}(k) / 7 ) .
\label{eq:cosweekday}
\end{equation}

In addition to the temporal features, two features related to past water draws improved predictive accuracy. Both features involve a tunable water mass flow threshold $\tau$ (kg/s). The first feature,
\begin{equation}
| \{ j \in \{1, \dots, 12\} \mid \dot m(k - j) \geq \tau \} | ,
\end{equation}
is the number of water draws over the past hour (12 time steps) that exceeded $\tau$. The second feature, 
\begin{equation}
\min \{ j \in \{1, \dots, 12\} \mid \dot m(k - j) \geq \tau \} ,
\end{equation}
is the number of time steps since the last water draw exceeded $\tau$. The importance of these existing and newly engineered features is subsequently analyzed using Shapley values, which quantify the contribution of each feature to the model output \cite{winter2002shapley}. Lagged features and rolling means of past measurements were also considered, but degraded rather than enhanced the forecasting performance.

\begin{figure*}[ht!]
\centering
\includegraphics[width=0.95\textwidth]{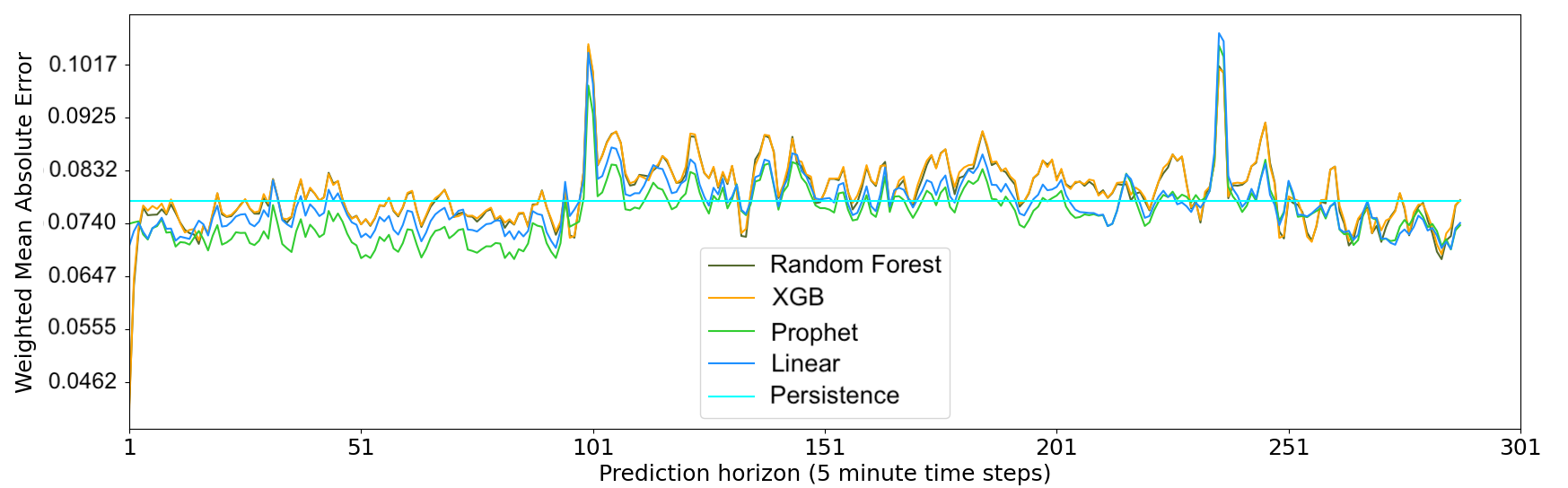}
\caption{Weighted mean absolute error over different prediction horizons for each forecast algorithm.}\label{fig:longhorizon}
\end{figure*}

\subsubsection{Model selection} 

The performance of several benchmark and state-of-the-art time series forecasting models is compared within the validation period. In the following, we briefly introduce the investigated models: 

\begin{itemize}
    \item \textbf{Persistence model:} One of the simplest time series forecasting models is a persistence model that takes the observations from the period before as forecast for the next period \cite{ghilardi2023benefits, lazos2015development}. This method is often used as a benchmark model, because of its easy implementation and reproducibility. 

    \item \textbf{Linear Regression:} The hot water draw forecast can be represented in a tabular format, where each set of features $\chi(k)$ at time step $k$ is mapped to a future observation $\dot m(k+j)$, with $j \in \{1, ..., K\}$, where $K$ is the forecasting horizon of 288 five-minute time steps. A linear regression model can then be trained to forecast all future observations $\dot m(k+j)$ for each $j$, by fitting a linear equation based on past observations. Linear regression offers low computational complexity and has been successfully applied in several time series forecasting studies \cite{saber2017short, xu2021automl}.
    
    \item \textbf{Random Forest:} The Random Forest algorithm, introduced by \cite{breiman2001random}, is itself an ensemble model of multiple decision trees, that can be used in tabular forecasting tasks. The Random Forest model exhibits good predictive performance, is prone to overfitting problems, and has a relatively low computational complexity \cite{grinsztajn2022tree, breiman2001random}. The previously described approach, to transform tabular data for time-series forecasting problems, is used for the Random Forest regression to forecast the hot water draw day-ahead time series. 

    \item \textbf{XGBoost:} The XGBoost model also employs an ensemble of sequentially built decision trees, where additional trees are built to correct the error of previously built trees with gradient tree boosting \cite{chen2016xgboost}. The method has been successfully applied for various time series forecasting tasks and comes with a high computational efficiency \cite{abbasi2019short, zhang2021time}.
    
    \item \textbf{Prophet:} The Prophet model is an easy-to-use time series forecasting model that decomposes the time series into various components. The trend of the time series is modeled with a saturating growth and piecewise linear model, while the seasonality is covered with a Fourier series \cite{taylor2018forecasting}. Shen et al. used the Prophet forecasting model in an MPC system that scheduled electric water heating for a multi-family residential building \cite{shen2021data}.
    
\end{itemize}

In initial experiments, alternative neural network-based models, such as the Temporal Fusion Transformer \cite{zhou2021informer}, were also investigated. However, these models were not pursued further due to their high computational costs and fitting times, which approached the control system's five-minute time step. Similarly, the ARIMA method was excluded because of its high runtime \cite{amin2019automating}. No hyper-parameter tuning was conducted, though this remains a potential avenue for future work.

\subsubsection{Evaluation} 

The presented models are evaluated using common forecasting metrics, including the Mean Absolute Error (MAE) and Root Mean Squared Error (RMSE) \cite{vom2020data}. In addition, the Weighted Mean Absolute Error (WMAE) is considered \cite{cleger2012use}. The WMAE is applied to assign greater importance to forecasts during actual hot water draw events. Considering the WMAE metrics is especially insightful, given that the target variable, the mean hot water draw over five minutes, is often zero due to the stochastic nature of showers and other hot water uses. Relying solely on traditional forecast metrics could favor forecasting techniques that predominantly predict zero, which is not suitable for this MPC application. To calculate the WMAE, the weighted sum of absolute errors between the actual values and the forecasted values, weighted by the actual values, is divided by the number of samples.

In addition to considering the three previously mentioned evaluation metrics, our study also considers the forecasting performance per method over varying prediction horizons. While most studies evaluate forecasts using an averaged value per metric, we consider this as too simplistic for effectively assessing forecasting models for MPC. This is particularly relevant when some models excel in short-term forecasts, while others perform better in the long-term, a phenomenon that has also been observed in other domains \cite{lazos2015development}. Averaging metrics alone can obscure these nuances. Therefore, it is suggested to compare averaged metrics across forecasting horizons, enabling the selection of forecasting models based on a detailed, horizon-dependent analysis. This approach ultimately informs the development of a hybrid predictive model.

\subsubsection{Ensemble model} 

As the individual models' forecasting performances were found to vary with the forecasting horizon, an ensemble model that utilizes different models for determined time horizons was built. At time $k$, the $j$-step-ahead forecast $\hat{\dot m}(j|k)$ of the future hot water mass flow $\dot m(k+j)$ depends on the current feature set $\chi(k)$, the lookahead $j$, the selected models $\Psi^{\text{short}}$, $\Psi^{\text{medium}}$, and $\Psi^{\text{long}}$, and the lookahead thresholds $J_1 < J_2 < J$:

\begin{equation}
\hat{\dot m}(j|k) =\begin{cases}
\Psi^{\text{short}}(\chi(k)) &\text{if } 1 \leq j < J_1 \\ 
\Psi^{\text{medium}}(\chi(k)) &\text{if } J_1 \leq j < J_2 \\  
\Psi^{\text{long}}(\chi(k)) &\text{if } J_2 \leq j \leq J .
\end{cases}
\label{eq:hybridmodel}
\end{equation}
Equation \ref{eq:hybridmodel} illustrates an ensemble model comprising three different models. In general, the appropriate number of models and thresholds can vary with the specific dataset and task. The model is calibrated based on the forecast validation dataset as detailed in Section \ref{sec:ForecastCaseStudy} for the field test site.

\subsection{Optimization formulation}
\label{sec:ControllerDesign}

Since the water heater's temperature dynamics are modeled as linear, a convex optimization problem was solved at each time step of the MPC algorithm. The aim is to minimize the operating costs of the HPOWH while maintaining comfortable outlet temperatures and respecting equipment constraints.

\subsubsection{Decision variables and control action}

In the MPC optimization problem solved at time $k$, the decision variables are the planned upper-node temperatures $T_u(0|k)$, \dots, $T_u(J|k)$, lower-node temperatures $T_\ell(0|k)$, \dots, $T_\ell(J|k)$, heat pump thermal power outputs $q(0|k)$, \dots, $q(J-1|k)$, and temperature set-points $T_s(0|k)$, \dots, $T_s(J-1|k)$. The notation $T_u(j|k)$ indicates the planned version of the future upper-node temperature $T_u(k+j)$, planned $j$ steps ahead at time $k$. The first planned temperature set-point $T_s(0|k)$ is implemented as the control action at time $k$. The system then evolves for five minutes, new forecasts are made, an updated optimization problem is solved, and the next control action $T_s(0|k+1)$ is implemented at time $k+1$. To match the water heater’s temperature set-point quantization, the implemented control actions are rounded to the nearest integer $^\circ$F (0.56  $^\circ$C).

\subsubsection{Objectives}

The objective function combines terms related to electricity costs, thermal comfort, and bacteria growth. The electricity cost,
\begin{equation}
\frac{\Delta t}{\eta} \sum_{j=0}^{J-1} c_\text{elec}(k+j) q(j|k) ,
\label{eq:objective_energy}
\end{equation}
involves a potentially time-varying electricity price $c_\text{elec}$ (\$/kWh) and the input electrical power $q(j|k) / \eta$ (kW) that is planned $j$ steps ahead at time step $k$. 

The thermal comfort penalty,
\begin{equation}
\gamma \Delta t \sum_{j=1}^J ( T_\text{min} - T_u(j|k) )_+ ,
\label{eq:objective_comfort}
\end{equation}
applies when the planned upper tank temperature $T_u(j|k)$ drops below a minimum comfortable threshold $T_\text{min}=37.7$ $^\circ$C \cite{showerTemp2}. Here $(\cdot)_+$ denotes the positive part, $\max(0, \cdot)$, and $\gamma$ is a tunable parameter. To prioritize comfort, the weight $\gamma$ (\$/ $^\circ$C/h) was set to 10 times the maximum electricity price $c_\text{elec}$ (\$/kWh) within the horizon.  

The bacteria growth penalty,
\begin{equation}
\gamma \Delta t \sum_{j=1}^J \pi(j|k) \left( T_\text{bact} - \frac{T_u(j|k)+T_\ell(j|k)}{2} \right)_+ ,
\label{eq:objective_bacteria}
\end{equation}
applies when the planned average tank temperature drops below a threshold $T_\text{bact} = 48.8$  $^\circ$C. Storing water at a long-term average temperature below $T_\text{bact}$ increases the risk of legionella bacteria growth \cite{Legionella_Temp}. After trial and error on the hardware, we found that temporarily disabling the bacteria growth penalty during and immediately after medium or large water draws improved the control system performance. This modification prevented the controller from overheating the tank in an attempt to maintain $(T_u + T_\ell)/2 > T_\text{bact}$ throughout an extended draw (e.g., a long shower). We implemented this modification through the time-varying, binary parameter

\begin{equation}
\pi(j|k) =
\begin{cases}
0 & \text{if } \exists i \leq 4 \text{ such that } \hat{\dot m}(j-i|k) > \phi \\
& \text{or } \Delta t \sum_{i=j-24}^{j} \hat{\dot m}(i|k) > \Phi \\
1 & \text{otherwise}.
\end{cases}
\label{eq:adaptive_objective}
\end{equation}
If any of the water flow rate forecasts $\hat{\dot m}(j-4|k)$, \dots, $\hat{\dot m}(j|k)$ exceed the threshold $\phi = 0.75$ kg/min, then $\pi(j|k) = 0$ and the bacteria growth penalty is inactive at time $k+j$. The relatively low threshold of $\phi$ was chosen to capture peaks predicted by the Prophet forecasting model, which tends to underestimate water draws; however, this value can be adjusted for other forecasting methods. Additionally, if the rolling two-hour (24 time steps) integrated hot water draw is predicted to exceed $\Phi = 18$ kg, then $\pi(j|k) = 0$ and the penalty is again inactive at time $k+j$. This case signals that a medium-usage event has occurred recently. Conversely, if neither of these conditions are forecasted to be met at time $k+j$, then $\pi(j|k) = 1$ and the bacteria growth penalty is active.

\subsubsection{Constraints}

In the optimization problem at time $k$, the planned node temperatures are initialized at the current measured node temperatures via equality constraints:
\begin{equation}
T_u(0|k) = T_u(k), \ T_\ell(0|k) = T_\ell(k) . \label{eq:initial_state}
\end{equation}
The forecasted dynamics,
\begin{equation}
\begin{aligned}
\begin{bmatrix} T_u(j+1|k) \\ T_\ell(j+1|k) \end{bmatrix} &= A(j|k) \begin{bmatrix} T_u(j|k) \\ T_\ell(j|k) \end{bmatrix} \\
&\quad + B(j|k) q(j|k) + w(j|k) , \label{eq:dynamics} 
\end{aligned}
\end{equation}
also enter as equality constraints. The matrices $A(j|k)$ and $B(j|k)$, as well as the disturbances $w(j|k)$, are constructed from the forecasted water draws $\hat{\dot m}(j|k)$ using the equations in Section \ref{sec:Modeling}. Forecasted ambient air temperatures and inlet water temperatures, $\hat{T}_a$ and $\hat{T}_c$, also appear in $w(j|k)$. As the water heater resided in an unconditioned basement, $\hat{T}_a$ was set to 1.7 $^\circ$C below the daily average space conditioning thermostat set-point. This logic can be tuned depending on where the water heater resides. The inlet water temperature $\hat{T}_c$ was assumed constant over the prediction horizon at the lowest value of the measured $T_c$ during a large draw, which better approximated the true municipal supply temperature.

In addition to the thermal dynamics in Equation \eqref{eq:dynamics}, the upper-node temperature $T_u$ is assumed to track the temperature set-point $T_s$ with first-order linear error dynamics. The discrete-time tracking dynamics,
\begin{equation}
T_u(j+1|k) = a T_u(j|k) + (1-a) T_s(j|k) , \label{eq:tracking_dynamics}
\end{equation}
show that the upper-node temperature is modeled as a mixture of the previous upper-node temperature and the previous temperature set-point. This mixture is weighted by the parameter $a = 0.8$, which was tuned to the tracking behavior observed in step-response experiments on the hardware. The temperature set-point is constrained below a maximum set-point $T_{s,\text{max}} = 60$  $^\circ$C:
\begin{equation}
T_s(j|k) \leq T_{s,\text{max}} . \label{eq:set-point_limit}
\end{equation}
This is the highest set-point allowed by the water heater's API. The API also has a minimum allowable set-point, but this is not enforced in the optimization problem to avoid infeasibility issues arising from the simplified tracking model. Instead, the lower limit on $T_s(0|k)$ is applied in a post-processing step, ensuring the implemented action lies within $[43.3,60]$  $^\circ$C ($[110, 140]$ $^\circ$F). 

The optimization problem at time $k$ also constrains the heat pump's thermal power output,
\begin{equation}
0 \leq q(j|k) \leq \eta P_\text{max} , \label{eq:capacity}
\end{equation}
where $P_\text{max}$ is the maximum electrical power input.

\subsubsection{Summary}

\begin{figure*}
\centering
\includegraphics[width=0.9\textwidth]{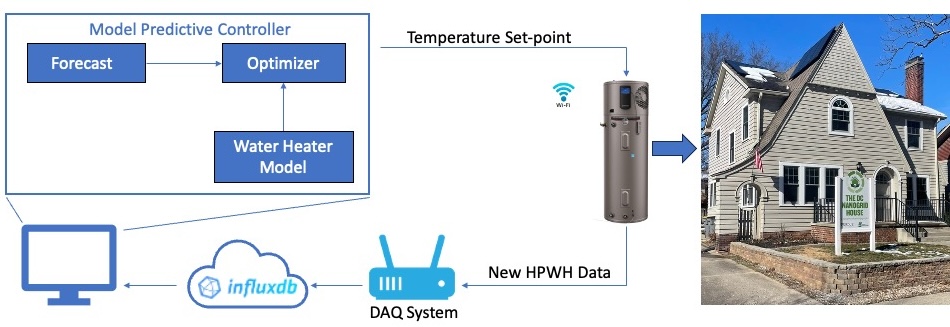}
\caption{IoT infrastructure in the test house. A data acquisition system passes sensor measurements to InfluxDB for cloud storage. A computer in the house runs the MPC and Python scripts that push the set-point adjustment to the water heater via API. The optimizer uses the hot water forecast to compute a new set-point at each time step.}\label{fig:DC House & IoT}
\end{figure*}

In summary, the MPC algorithm solves the following optimization problem at each time $k$.

\begin{equation}
\label{eq:full_formulation}
\begin{aligned}
&\text{Minimize Eq.} \ \eqref{eq:objective_energy} + \text{Eq. } \eqref{eq:objective_comfort} + \text{Eq. } \eqref{eq:objective_bacteria} \\
&\text{Subject to}\\
&\quad \text{Initial state: Eq. } \eqref{eq:initial_state}  \\ 
&\quad \text{Thermal dynamics: Eq. } \eqref{eq:dynamics} \text{ for } j = 0, \ \dots, \ J-1 \\
&\quad \text{Set-point tracking: Eq. } \eqref{eq:tracking_dynamics} \text{ for } j = 0, \ \dots, \ J-1 \\
&\quad \text{Set-point limits: Eq. } \eqref{eq:set-point_limit} \text{ for } j = 0, \ \dots, \ J-1 \\
&\quad \text{Heat pump capacity: Eq. } \eqref{eq:capacity} \text{ for } j = 0, \ \dots, \ J-1 \\
\end{aligned}
\end{equation}

The decision variables are the planned trajectories of the upper- and lower-node temperatures, heat pump thermal power, and temperature set-point. With a prediction horizon of one day ($J = 288$ five-minute time steps), this problem has 1,174 decision variables, 578 linear equality constraints, and 864 linear inequality constraints. As the objective function is piecewise linear in the decision variables, the problem can be converted to a standard-form linear program and solved efficiently using off-the-shelf simplex or interior-point methods. In our field experiments, the optimization problem was solved in Matlab \cite{MATLAB} using the Gurobi \cite{gurobi} solver and the CVX modeling language \cite{cvx}. Solving one instance of this problem took about 0.06 seconds on a 5.6 GHz processor with 32 GB of system memory. Solving the MPC problem \eqref{eq:full_formulation} is likely efficient enough to enable implementation on a low-cost microcontroller.

\section{Test house}
\label{sec:Test House}

The field demonstrations in this paper took place in a test house, pictured in Figure \ref{fig:DC House & IoT}, near Purdue University's campus in West Lafayette, Indiana, USA. The test house is a 208 m$^2$, two-story, 1920s-era detached single-family home. Referred to as the DC Nanogrid House \cite{CHPB_DCNanogridHouse}, the test house has both AC and DC electrical systems. For the demonstrations in this paper, no changes were made to the existing equipment and everything ran on AC power. The home has three year-round occupants. The hot water in the house is produced via a 189.3 L (50-gallon) hybrid-HPWH equipped with two resistance heating elements \cite{RheemHPWH}. The water heater has a Uniform Energy Factor of 3.75 \cite{uefrating}. The upper resistance heating element is 2.25 kW and the lower element is 4.5 kW. The water heater is also equipped with two thermistors on the upper and lower parts of the tank used in the default controls. The water heater has four different operational modes: resistance-only mode, high demand mode, power savings mode (hybrid mode), and heat pump-only mode.

\begin{figure}
\centering
\includegraphics[width=0.42\linewidth]{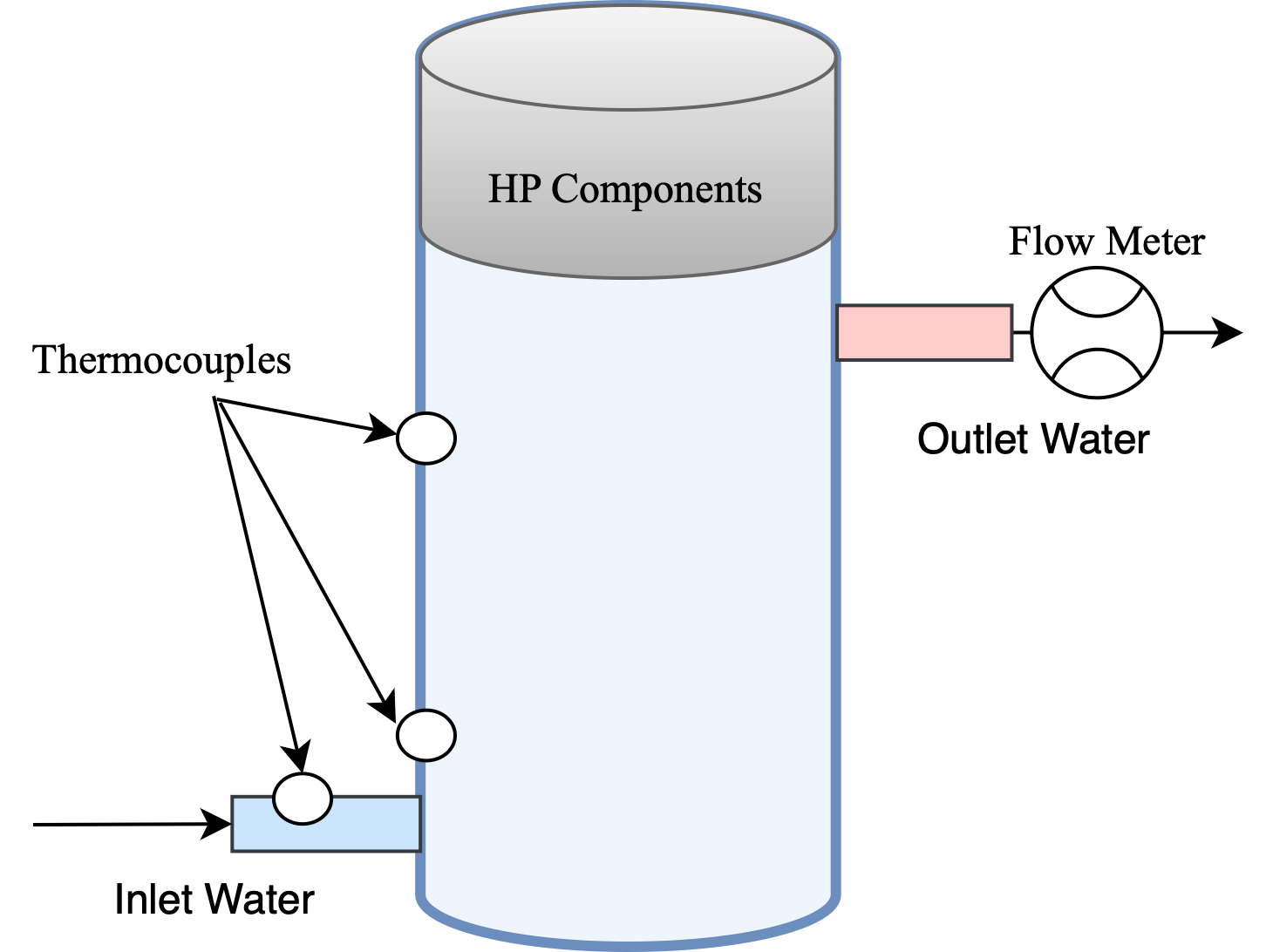}
\caption{Sensors on the HPOWH include thermocouples on the upper and lower tank exterior and the inlet line, as well as a flow meter on the outlet line.}
\label{fig:sensing_equip}
\end{figure}

During typical operation, the water heater runs in hybrid mode to ensure comfort. This mode turns on the resistance heating elements before water temperatures become uncomfortably low. Running the water heater in heat-pump-only mode reduces the occupants' energy costs but increases the risk of discomfort. The compressor size is often the same or similar for 50-gallon HPOWHs and hybrid-HPWHs in the market \cite{RheemHPWH}. Therefore, when running the water heater used in the field test in heat-pump-only mode, it resembles a typical HPOWH. 

This paper aims to develop a low-cost control system that requires minimal instrumentation. Therefore, minimal sensing equipment was added to ensure that the control system had the necessary measurements for the water heater. The water heater's API enables adjustment of set-point temperatures but does not allow remote readings from the manufacturer's thermistors. For this reason, two thermocouples were placed on the upper and lower parts of the tank exterior, next to the manufacturers' thermistors. This set-up mirrors the measurements that would be available to the manufacturer or from other models that make these measurements available via API \cite{Starke2020}. The sensor on the upper part of the tank is also used to estimate the outlet water temperatures. The only additional sensing equipment that is not used by the manufacturer is a flow sensor on the hot water line and a thermocouple on the inlet water line to the tank (Figure \ref{fig:sensing_equip}). The flow sensor provides accurate hot water draw measurements that enable predictions of future patterns. The thermocouple on the inlet line is used to estimate the heat loss due to cold inlet water balancing hot water draws. The data acquisition system interprets these measurements, which are fed by a RaspberryPi to the cloud database, InfluxDB \cite{influxdb}, using Python scripts \cite{pergantis2024field}. The controller is then operated on a computer in the home that pushes the set-point adjustments to the water heater, as shown in Figure \ref{fig:DC House & IoT}. The IoT architecture described above could be implemented on a single-board computer that acts as the data acquisition system, stores the minimum necessary data, and runs MPC to update the water heater set-point. To enable the MPC algorithm to run with limited computing power, the computationally inexpensive linear convex formulation for the optimization algorithm in the the MPC was chosen.

\subsection{Comfort issues with HPOWHs}

To explore the comfort performance of a HPOWH, the water heater was switched to heat-pump-only mode for a trial period. During this period, the set-point was held constant at the historical mean value of 48.9  $^\circ$C (120 $^\circ$F). The test ran for five days (April 6--10, 2025) but was stopped due to occupant complaints of showers being cut short by cold water. An outlet temperature is considered `cold' if it is below 37.7 $^\circ$C (100 $^\circ$F), the temperature at which showers typically become uncomfortable \cite{LUOComfort2023}.

\begin{figure}
\centering
\includegraphics[width=0.5\linewidth]{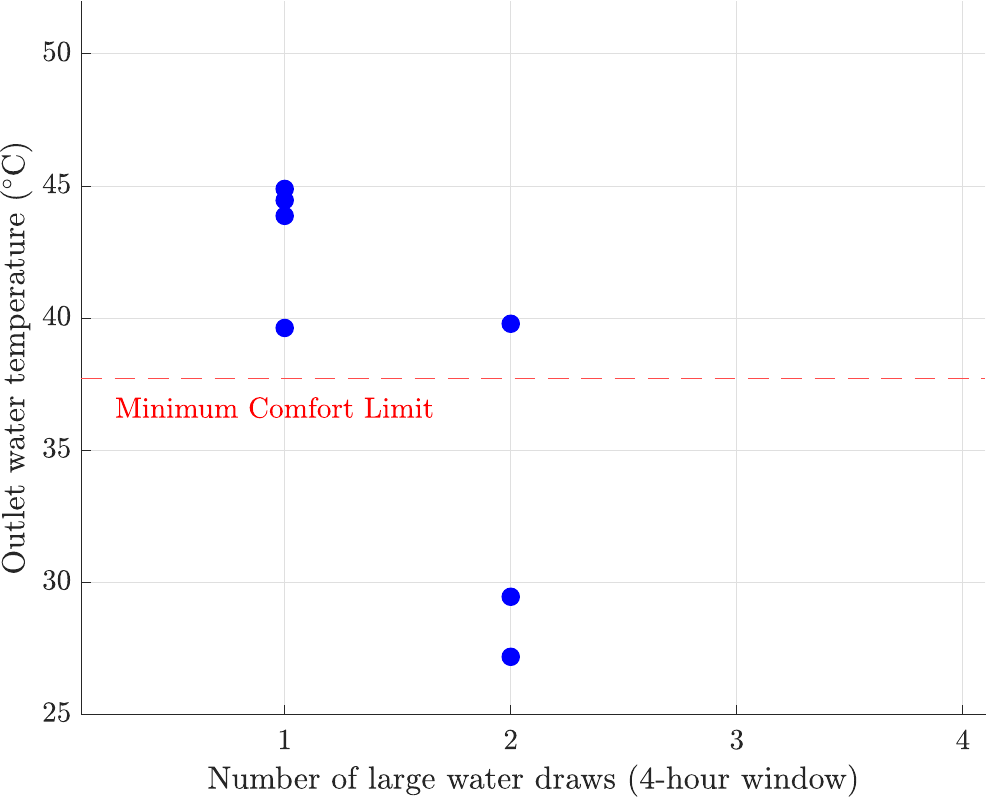}
\caption{Lowest outlet water temperature during large water draws over a four-hour window vs. number of large draws in the same window. {\color{blue} Default HPOWH controls} often cause discomfort during consecutive showers.}\label{fig:Temps Baseline}
\end{figure}

To find the root causes of the cold temperatures, hot water draw data was analyzed, focusing on large water draws, defined as any event consuming more than 18.9 L (5 gallons). The number of these events were aggregated over a four-hour window. A four-hour window was used to capture the morning period in which the occupants often take showers. The lowest outlet temperature during water draws within this window was then plotted in Figure \ref{fig:Temps Baseline}. The analysis revealed that cold outlet temperatures occurred due to consecutive large water draws within the same window.

Over the five-day trial period, there were 10 large water events, nine of which were showers. The outlet temperature often dropped below the comfort limit when two showers were taken within a single four-hour window. The second shower then experienced cold outlet temperatures, seen by the lowest temperature data point in Figure \ref{fig:Temps Baseline}. With the manufacturers default controls and the resistance heating elements disabled, the heat pump's heat transfer output was insufficient to heat the incoming water. In hybrid-HPWH mode, one of the resistance heating elements would have turned on in these scenarios to supply a faster heat transfer rate to the water. The only instance where two consecutive water draw events met minimum comfort was when a shower occurred at the start of the four-hour window and a washing machine ran toward the end.

The significant discomfort observed with default manufacturer controls in HPOWH mode motivated the predictive control development in this paper. The controller developed here uses predictions of hot water use to strategically preheat in order to maintain thermal comfort. Alternatively, one could simply raise the HPOWH set-point temperature. However, this would significantly reduce the COP, increase electricity use and would not unlock the load-shifting capabilities that predictive control provides \cite{BAUMANN2023112923, DELAROSA2025_Rule}.

\begin{figure}
\centering
\includegraphics[width=0.5\linewidth]{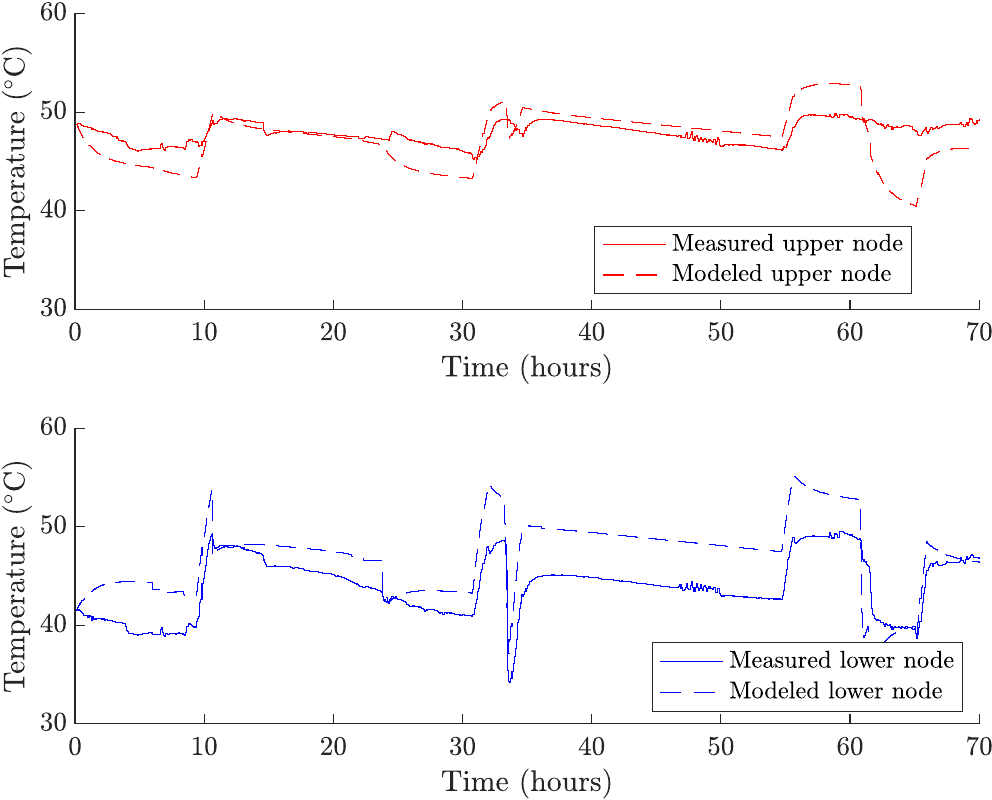}
\caption{Measured and modeled upper- and lower-node temperatures. The upper node model agrees well with measurements. The lower node model is less accurate but remains reasonably consistent with measurements.}\label{fig:modeled_dynamics}
\end{figure}

\subsection{Applying the methodology}

This section describes how the methodology in Section \ref{sec:methodology} was applied to the field test house. The water heater model was tuned from the field test house data and the water draw prediction algorithm was designed for the home.

\subsubsection{Water heater modeling}
\label{sec:Modeling case study}

The model parameters were determined based on manufacturer specifications and empirical data. The thermal capacitance of the tank was set to $C = 0.197$ kWh/$^\circ$C, derived from the water heater's volume as specified in the product datasheet. The thermal resistance of the tank insulation was set to $R_a = 1476$ $^\circ$C/kW, also based on the insulation properties provided in the datasheet.

To tune the other model parameters -- the COP $\eta$, the stratification height $z$, the stratification layer thickness $h_s$, and the fraction $\lambda$ of the heat pump's thermal power output that enters the upper node -- we used the methods in Section \ref{sec: ParameterTuning} and data recorded at one-minute intervals over representative summer and winter weeks. The tuned values were $\eta = 3.5$, $z = 0.5$, $h_s = 0.025$ m, and $\lambda = 0.3$.

Figure \ref{fig:modeled_dynamics} shows the measured (solid curves) and modeled (dashed) temperature dynamics of the upper (top plot) and lower (bottom) tank nodes. The upper plot shows good agreement between the measured and modeled upper-node temperatures. The lower node model, while somewhat less accurate, still reproduces the general trends of the measured dynamics. To create Figure \ref{fig:modeled_dynamics}, the model was initialized at $t = 0$ with the measured node temperatures, then run in open loop over a 70-hour horizon with five-minute time steps. This open-loop prediction task is more challenging than the MPC prediction task, which re-initializes the model with the current node temperature measurements at each time step. The open-loop prediction errors were assessed for both power consumption and node temperatures. Data from separate months in winter and summer were used for validation. The mean absolute power prediction errors were 2.1\% in summer and 3.9\%  in winter. The mean absolute upper- and \textcolor{black}{lower-node} temperature errors were 3.8\% \textcolor{black}{and 7.4\%} in summer and 8.1\% \textcolor{black}{and 13.7\%} in winter, respectively. \textcolor{black}{Colder winter inlet water produces more dynamic stratification: the lower node experiences faster heat transfer, increasing the temperature gap between nodes and amplifying error in the inter-node resistance $R_{u\ell}$, stratification height $z$, and the constant COP $\eta$. Similar dynamics also occur in summer during large hot water draws, when the lower-node temperature drops, though they are less pronounced because of the higher inlet water temperatures.} Although higher accuracy could be achieved with higher-order modeling, the additional sensing requirements and/or engineering effort would reduce the scalability of the algorithm. These results indicate that the reduced order model captures the essential thermal and electrical dynamics across different seasonal conditions.

\begin{figure}[h]
\centering
\begin{subfigure}[b]{0.48\linewidth}
    \includegraphics[width=\linewidth]{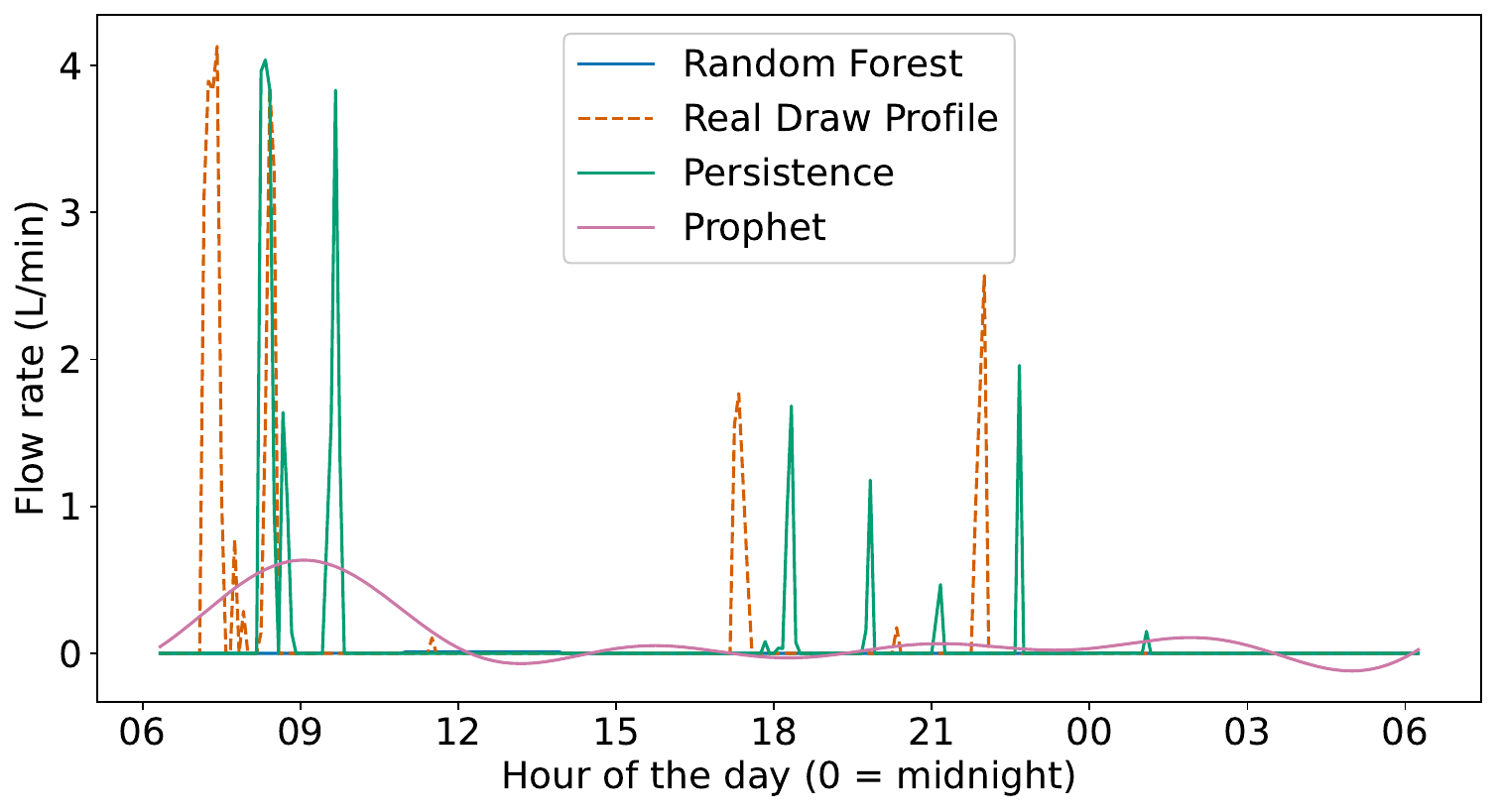}
    \caption{Before hot water draws.}
    \label{fig:beforedraw}
\end{subfigure}
\hfill
\begin{subfigure}[b]{0.48\linewidth}
    \includegraphics[width=\linewidth]{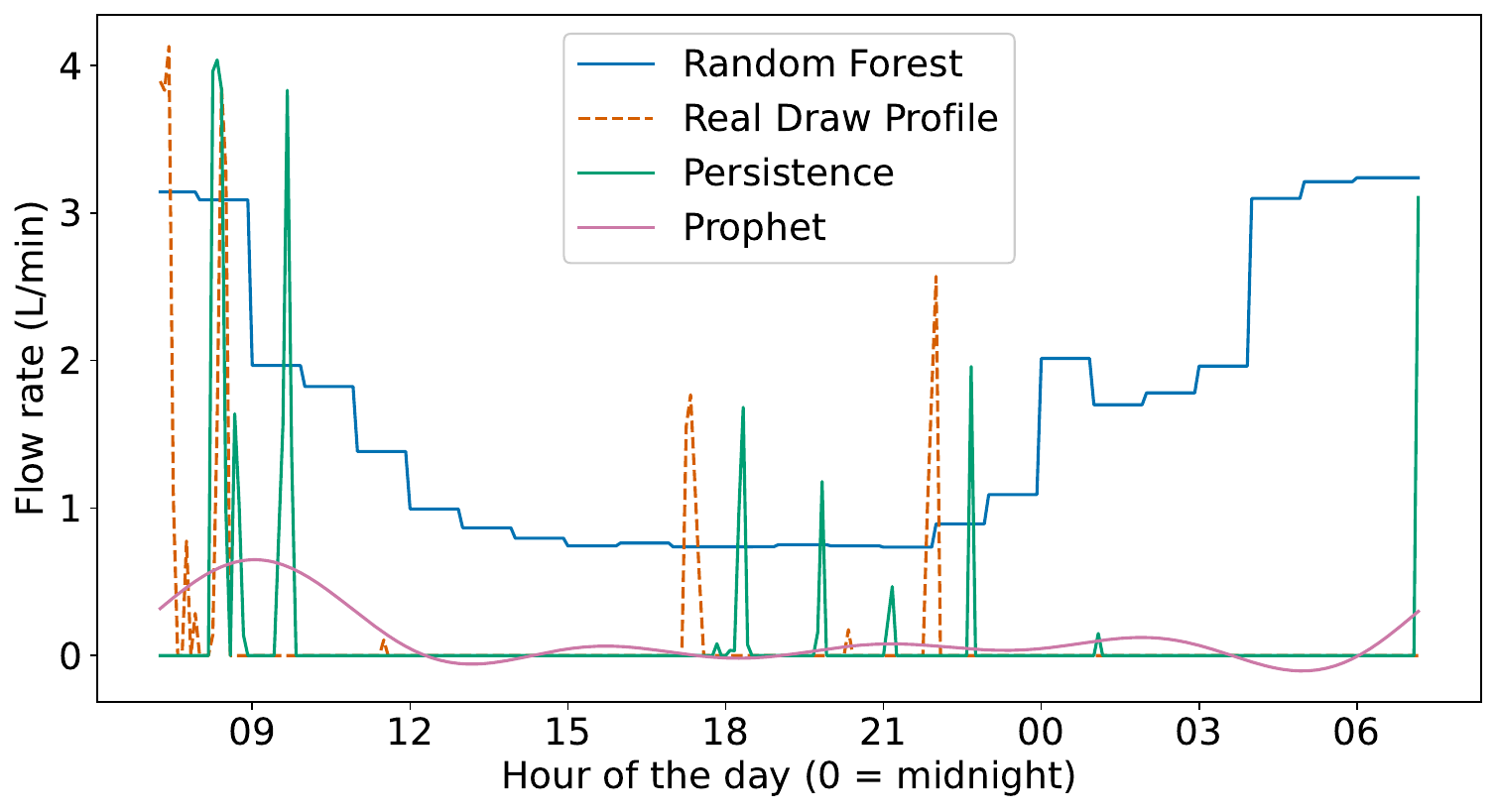}
    \caption{During hot water draw.}
    \label{fig:whiledraw}
\end{subfigure}
\hfill
\begin{subfigure}[b]{0.48\linewidth}
    \includegraphics[width=\linewidth]{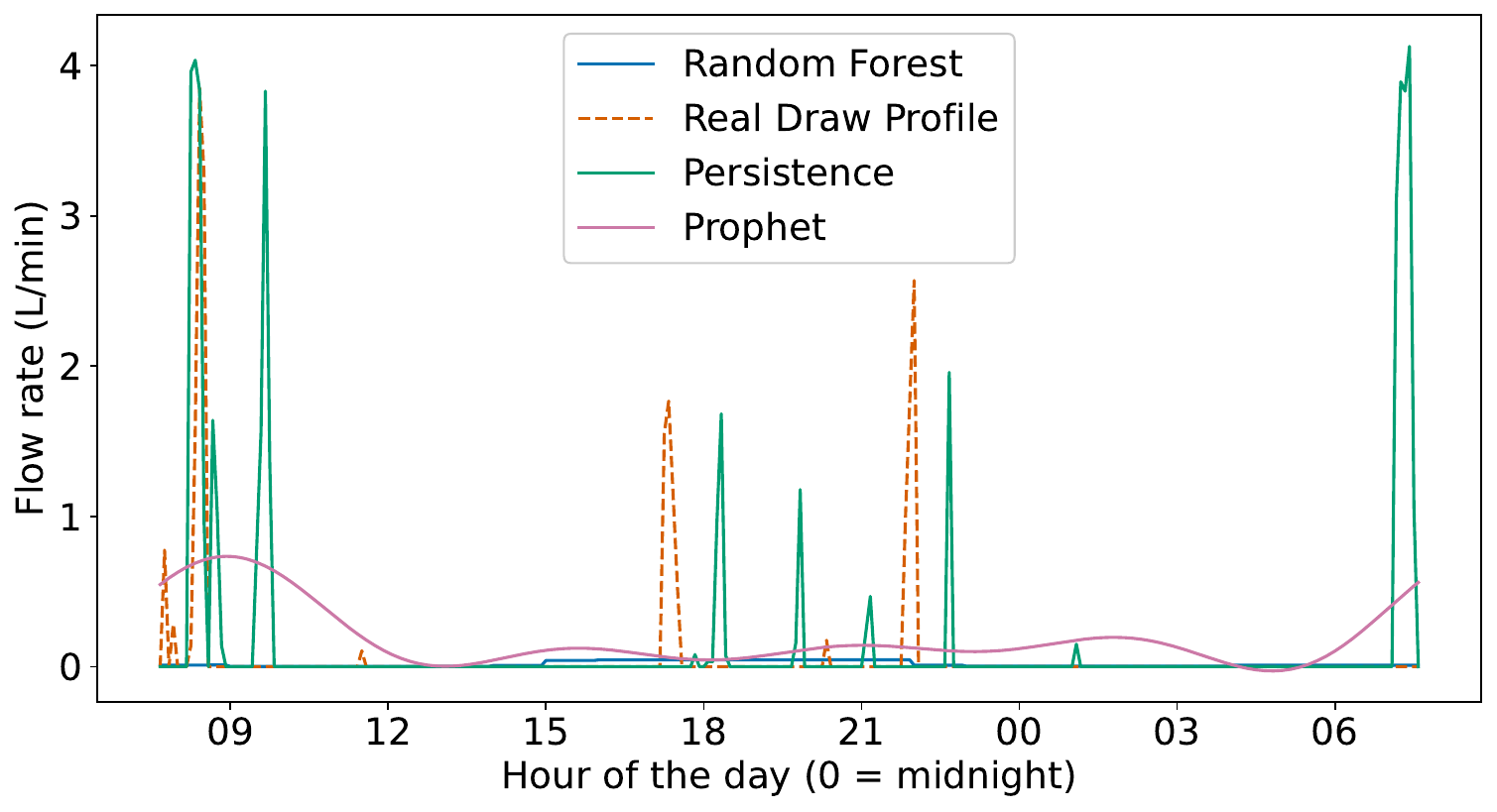}
    \caption{After hot water draw.}
    \label{fig:afterdraw}
\end{subfigure}
\caption{Forecasted behavior of the Random Forest, persistence, and Prophet models before, during, and after a water draw, compared against the actual draw profile.}
\label{fig:forecastdraw}
\end{figure}

\subsubsection{Water draw forecasting}
\label{sec:ForecastCaseStudy}

Measurements from March 10--17, 2024 were used to validate the forecasting models, with training data extending back to February 14, 2024. The models were re-trained every five minutes to generate forecasts for the subsequent 24 hours with five-minute time steps. Table \ref{table:forecastresults} presents the overall performance of the compared models across the specified metrics. The Prophet performed best with respect to all of the high-level metrics.

When examining the average WMAE across different prediction horizons in Figure \ref{fig:longhorizon}, a more nuanced view of forecasting performance over time emerges. For the first three time steps (15 minutes), the tree-based XGBoost and Random Forest models outperformed the other models, while the Prophet model performed particularly well for the first 100 time steps. Beyond this range, the persistence model performed best. To maximize overall accuracy, we used the Random Forest model to forecast $1$, \dots, $J_1 = 4$ steps ahead, the Prophet model to forecast $5$, \dots, $J_2 = 100$ steps ahead, and the persistence model to forecast $101$, \dots, $J = 288$ steps ahead. This ensemble approach contrasts with previous studies, which have often relied on a single model such as Prophet \cite{shen2021data}. In our case, using only the single model that performed best with respect to the high-level metrics (Prophet) would have worsened our short-term forecasts and MPC performance.

\begin{table*}
\centering
\caption{Forecasting model comparison based on RMSE, MAE, and WMAE.}
\begin{tabular}{lrrrrr}
 & Random Forest & Persistence & Prophet & XGBoost & Linear Regression \\
\midrule
RMSE (L) & 0.625 & 0.762 & 0.521 & 0.617 & 0.558 \\
MAE (L) & 0.222 & 0.218 & 0.214 & 0.218 & 0.259
 \\
WMAE (-) & 2.553 & 2.553 & 2.4272 & 2.5604 & 2.486 \\
\end{tabular}
\label{table:forecastresults}
\end{table*}

Figure \ref{fig:forecastdraw} shows the three selected models' forecasts before, during, and after a 7 AM shower. The Prophet model provides a balanced forecast, indicating potential showers between 6 AM and noon, with the time integral approximating typical cumulative morning water draw over the period. During the shower, both the Prophet and persistence forecasts fail to adapt, predicting continued hot water draw in subsequent time steps. In contrast, the Random Forest forecast reacts to the start of the shower, accurately forecasting a large water draw at the next time step. Shortly after the shower ends, the Random Forest model quickly adapts again, predicting no further showers that morning.

The trained Random Forest model, interpreted using Shapley values \cite{winter2002shapley}, reveals that features related to recent water draws play a dominant role in the predictions. Specifically, shorter durations since the last large water draw are strongly associated with higher predicted values in subsequent time steps. The frequency of recent water draws further contributes to the model’s output, reinforcing the relevance of short-term usage patterns. Temporal features also exhibit considerable influence, with cyclical representations of time (e.g., hour and weekday) demonstrating greater predictive value than their non-cyclical counterparts. These findings underscore the advantage of using cyclical time encoding for forecasting applications \cite{semmelmann2022load}.

\section{Results}
\label{sec:results}

The MPC algorithm was applied to the HPOWH under three electricity rate structures: flat, time-of-use (TOU), and hourly pricing. The flat rate structure, wherein the electricity price is time-invariant, serves as a baseline to compare the thermal comfort achieved under the MPC operated HPOWH and hybrid-HPWH. The TOU rate structure reflects a common two-tier pricing option offered by many utilities, with a higher peak price from 2--8 PM and a lower off-peak price at all other times. Under hourly pricing, the retail electricity price is dynamically adjusted each hour based on clearing prices in the wholesale day-ahead energy market. The TOU and day-ahead rate structures highlight the load shifting capabilities of the MPC algorithm, making the HPOWH demand response ready as well.

\subsection{Constant price case study}
\label{constantPrice}

A field test with a flat electricity rate from a local utility, Tipmont, was used to evaluate the thermal comfort provided by the MPC algorithm. The electricity rate was 0.1241 \$/kWh. This structure offers a straightforward baseline to isolate and assess the effects of the algorithm without the influence of varying electricity rate complexities. Testing occurred from May 4--6 and September 29--October 17, 2024, totaling 20 days. A summer hiatus interrupted testing, and two additional days in October were excluded due to power outages affecting IoT functionality. Results were compared against the five HPOWH baseline days (April 6--10) and a 24-day hybrid-HPWH baseline period (June 20--27, August 29--September 10, October 20--23). 

\begin{figure}
\centering
\includegraphics[width=0.5\linewidth]{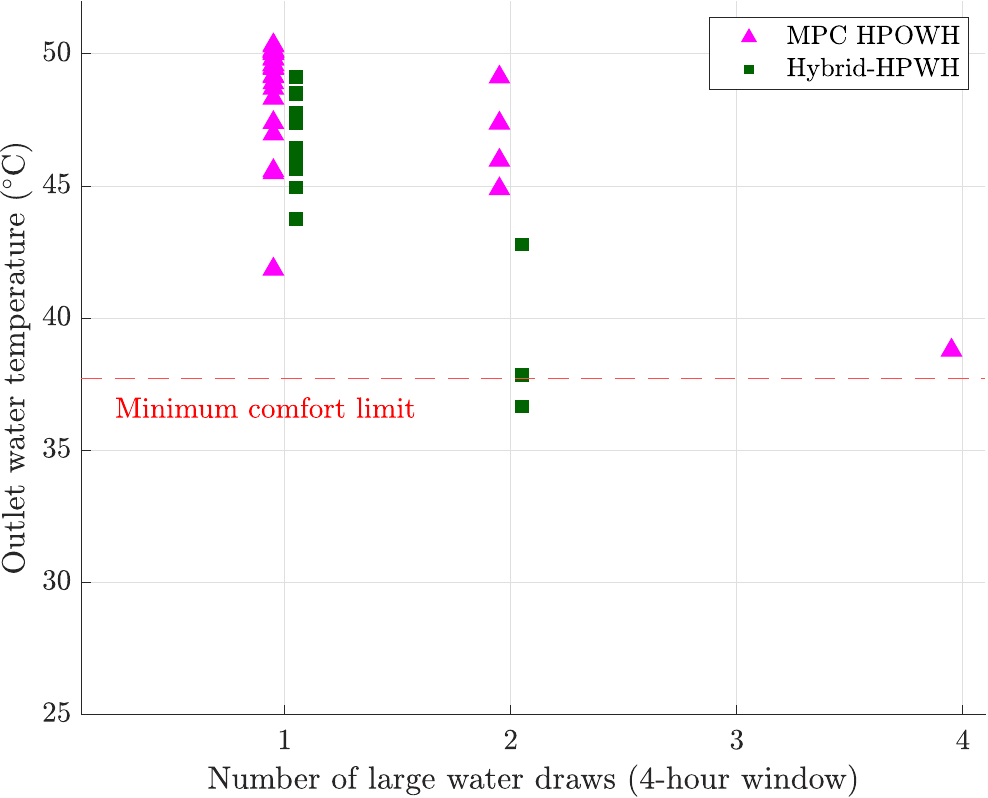}
\caption{Lowest outlet water temperature during large water draws over a four-hour window vs. number of large draws in the same window. The {\color{ForestGreen} hybrid-HPWH with default controls} mostly maintains comfort. The {\color{magenta} HPOWH with MPC} eliminates the discomfort caused by {\color{blue} HPOWH default controls} in Figure \ref{fig:Temps Baseline}.}
\label{fig:mpc_hybrid_comf}
\end{figure}

Extended MPC and hybrid-HPWH testing ensured fair comparisons without risking uncomfortable showers for occupants, unlike the baseline HPOWH testing. Peak water draw days were 256.5 L (HPOWH baseline), 287.2 L (hybrid-HPWH baseline), and 274.8 L (MPC). Water draw profiles across all modes were similar, with multiple large draw events often occurring within four-hour windows (Figures \ref{fig:Temps Baseline} and \ref{fig:mpc_hybrid_comf}).

Figure \ref{fig:mpc_hybrid_comf} shows hybrid-HPWH comfort performance. Unlike the HPOWH baseline, seen in Figure \ref{fig:Temps Baseline}, the hybrid-HPWH maintained comfortable outlet water temperatures during consecutive showers, with only two minor deviations. Occupants reported no discomfort during the hybrid-HPWH testing. The hybrid-HPWH maintained comfortable outlet water temperatures by using resistance heating elements, increasing energy use.

Figure \ref{fig:mpc_hybrid_comf} also shows that the HPOWH with MPC maintained comfortable outlet water temperatures, even on days with four consecutive large water draws within a four-hour window. This contrasts starkly with the unacceptable comfort performance of the HPOWH under default control, shown in Figure \ref{fig:Temps Baseline}. The comfort improvement came mainly from MPC preheating before large water draws.

\begin{figure}[h]
\centering
\begin{subfigure}[b]{0.98\linewidth}
    \includegraphics[width=\linewidth]{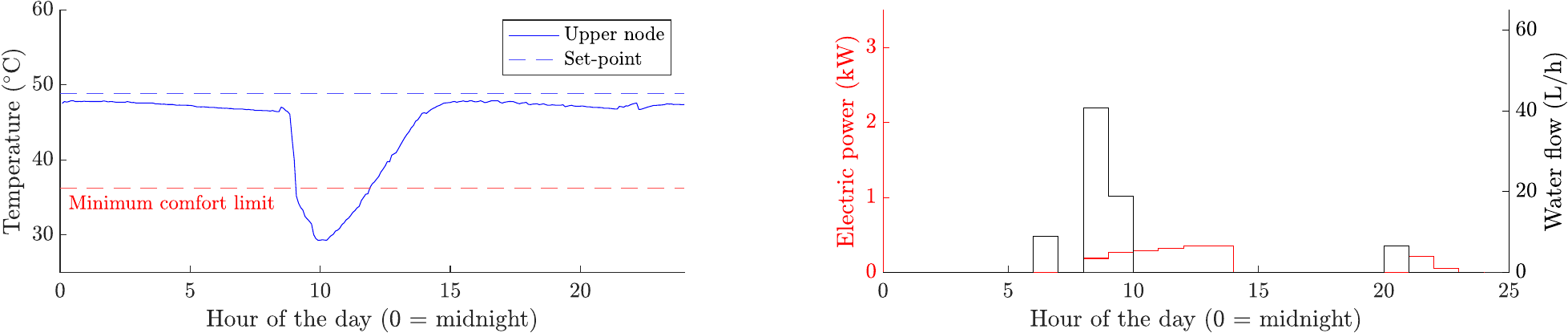}
    \caption{For the {\color{blue} HPOWH with default controls}, outlet water temperatures (left, solid blue) drop during large water draws (right, black) due to insufficient heat pump capacity (right, red).\newline}
    \label{fig:example_day_HPOWH}
\end{subfigure}
\hfill
\begin{subfigure}[b]{0.98\linewidth}
    \includegraphics[width=\linewidth]{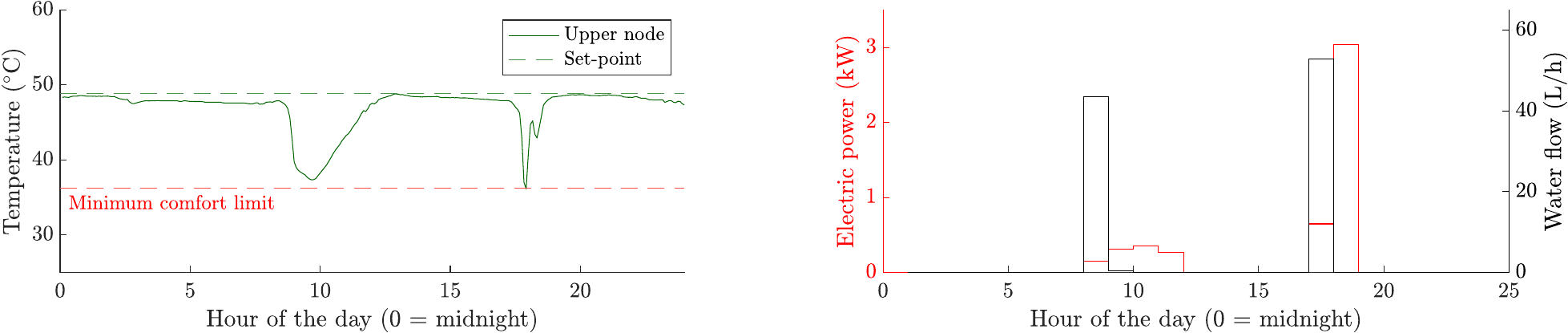}
    \caption{For the {\color{ForestGreen} hybrid-HPWH with default controls}, outlet water temperatures (left, solid green) remain comfortable during large water draws (right, black) due to additional heating capacity (right, red) from the (inefficient) heating elements.}
    \label{fig:example_day_hybrid-HPWH}
\end{subfigure}
\hfill
\begin{subfigure}[b]{0.98\linewidth}
    \includegraphics[width=\linewidth]{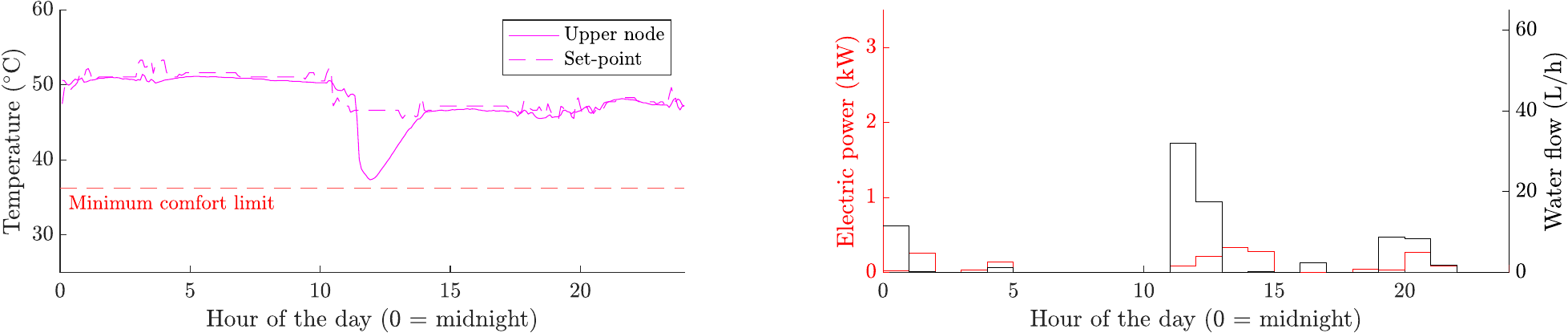}
    \caption{For the {\color{magenta} HPOWH with MPC}, outlet water temperatures (left, magenta) remain comfortable during large water draws (right, black) with much less energy use (right, red) than the {\color{ForestGreen} hybrid-HPWH with default controls} in Figure \ref{fig:example_day_hybrid-HPWH}.}
    \label{fig:example_day_MPC_flatrate}
\end{subfigure}
\caption{Comparison of outlet water temperature and control behavior in a flat rate structure across three operating modes: {\color{blue} HPOWH with default controls}, {\color{ForestGreen} hybrid-HPWH with default controls}, and {\color{magenta} HPOWH with MPC}.}
\label{fig:example_controls_comparison}
\end{figure}

Figures \ref{fig:example_day_HPOWH}, \ref{fig:example_day_hybrid-HPWH}, and \ref{fig:example_day_MPC_flatrate} show example days for the HPOWH with default controls, hybrid-HPWH, and HPOWH with MPC, respectively. In these figures, the top plots show the water temperature set-points, the upper-node temperatures, and the minimum comfort limit. The bottom plots show the hourly input electric power (left axes) and the water draw flow rate (right axes). 

Figure \ref{fig:example_day_HPOWH} shows that the HPOWH with default controls did not maintain comfortable outlet water temperatures. The upper-node temperature (a stand-in for the outlet temperature) dropped as much as 10 $^\circ$C below the minimum comfort limit. This behavior resulted from the limited heating capacity of the heat pump and the inability of default controls to preheat before large water draws. In hybrid-HPWH mode, the resistance heating elements would have turned on during large water draws, providing warmer water but using more energy.

Figure \ref{fig:example_day_hybrid-HPWH} shows that the hybrid-HPWH maintained comfortable outlet water temperatures by using its resistance heating elements. When the upper-node temperature dropped below approximately 37.7 $^\circ$C \textcolor{black}{around hour 18 of the day,} the 4.5 kW heating element turned on and the heat pump turned off. This transition heated water faster but used more energy, \textcolor{black}{underscoring the hybrid-HPWH's vulnerability to demand peaks that trigger inefficient resistance heating.} The heating element also requires a 240 V circuit.

Figure \ref{fig:example_day_MPC_flatrate} shows that the HPOWH with MPC dynamically varies the water temperature set-point to shift thermal load over time. MPC raises the set-point prior to the morning period, preheating in anticipation of large water draws. After the large water draws between 10 AM and noon, MPC lowers the set-point for the rest of the day, as further demand is predicted to be minimal and storing water at lower temperatures reduces thermal losses.

The MPC-operated HPOWH averaged 14.3 Wh/L, compared to 13.6 Wh/L for the hybrid-HPWH -- a small energy increase. Under the flat rate of 0.1241 \$/kWh, this translates to 0.00178 \$/L for the MPC mode versus 0.00169 \$/L for the hybrid configuration. The slightly higher energy use and cost in the MPC system are primarily attributed to its anticipatory control strategy, which preheats water based on forecasted usage patterns. When actual hot water demand is lower than expected, this preheating results in mild energy inefficiencies. Conversely, the hybrid-HPWH operates reactively, only heating when water temperature drops below thresholds defined by the manufacturer's control logic. Consequently, resistance heating elements are activated less frequently during typical demand, resulting in marginally lower energy use during this field test. These results may not generalize across all scenarios. In colder climates with lower inlet water temperatures, or in homes exhibiting different usage patterns -- such as higher peak demand or irregular draw schedules -- hybrid systems may rely more heavily on resistance heating, reducing efficiency and comfort. \textcolor{black}{In addition, for manufacturers who choose different HPOWH equipment configurations to increase the heating capacity, such as a larger compressor or a different refrigerant, the comfort problem may be less pronounced. However, a larger compressor would also change the electrical requirements for the HPOWH, as discussed in Section \ref{sec:discussion}, as well as the Uniform Energy Factor rating.}

While the MPC-operated HPOWH used slightly more energy and operating costs under current conditions, it offers potential monetary advantages when considering the infrastructure upgrades often required by hybrid-HPWH systems. These trade-offs are analyzed further in Section \ref{monthlyCosts}.

\subsection{Time-of-use pricing case study}

The second rate structure used was a two-tier TOU schedule from a local utility, Tipmont Rural Electric Membershop Cooperative \cite{tipmont2025tou}. The peak price was 0.251 \$/kWh from 2--8 PM every day. The off-peak price was 0.082 \$/kWh for all other hours of the day. The time-average electricity price for the day is equal to the time-invariant price from Section \ref{constantPrice}, enabling a fair cost comparison. MPC was tested with TOU prices from April 27-30, 2024.

\begin{figure}[h]
\centering
\begin{subfigure}[b]{0.48\linewidth}
    \includegraphics[trim={0.2cm 0cm 0.25cm 0cm},clip,width=\linewidth]{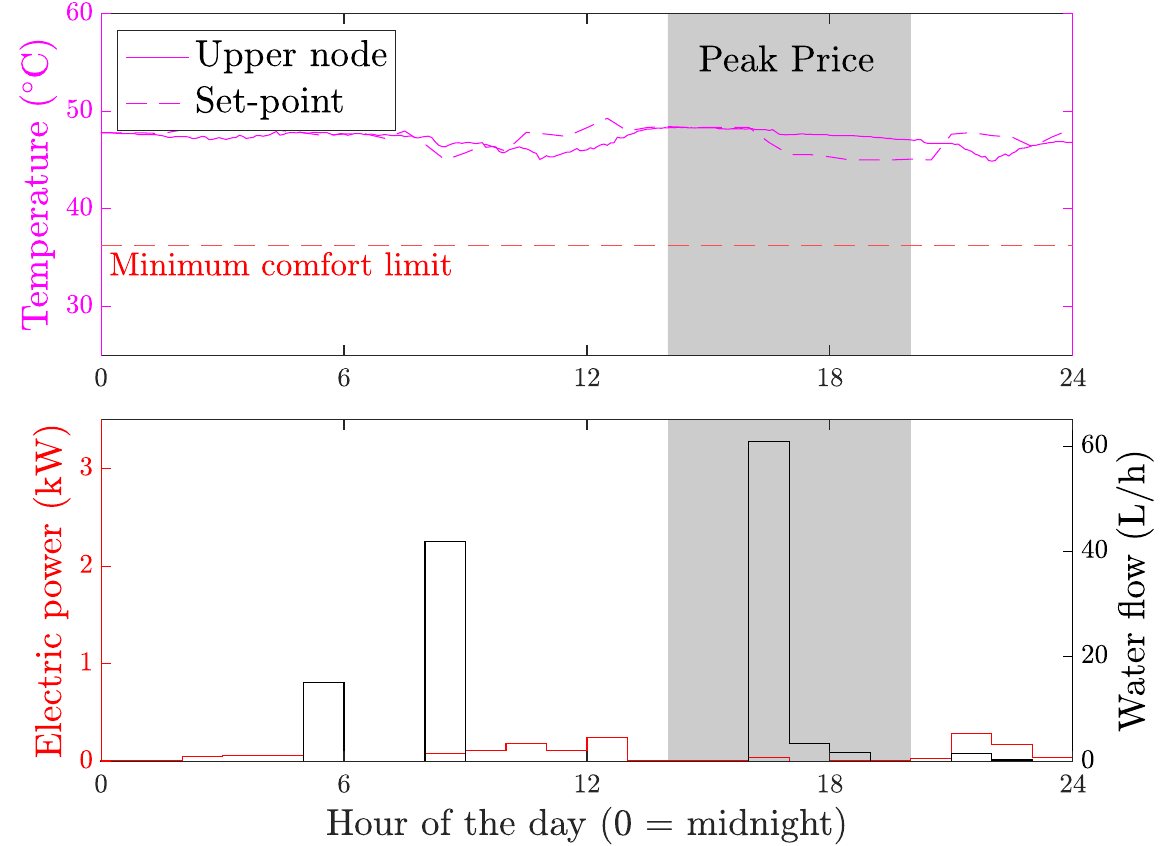}
    \caption{Under two-tier TOU pricing, the {\color{magenta} HPOWH with MPC} preheats before the peak price period (   gray), mostly drifts during, and reheats after. \newline \newline}
    \label{fig:tou_example}
\end{subfigure}
\hfill
\begin{subfigure}[b]{0.48\linewidth}
    \includegraphics[trim={0.2cm 0cm 0.25cm 0cm},clip,width=\linewidth]{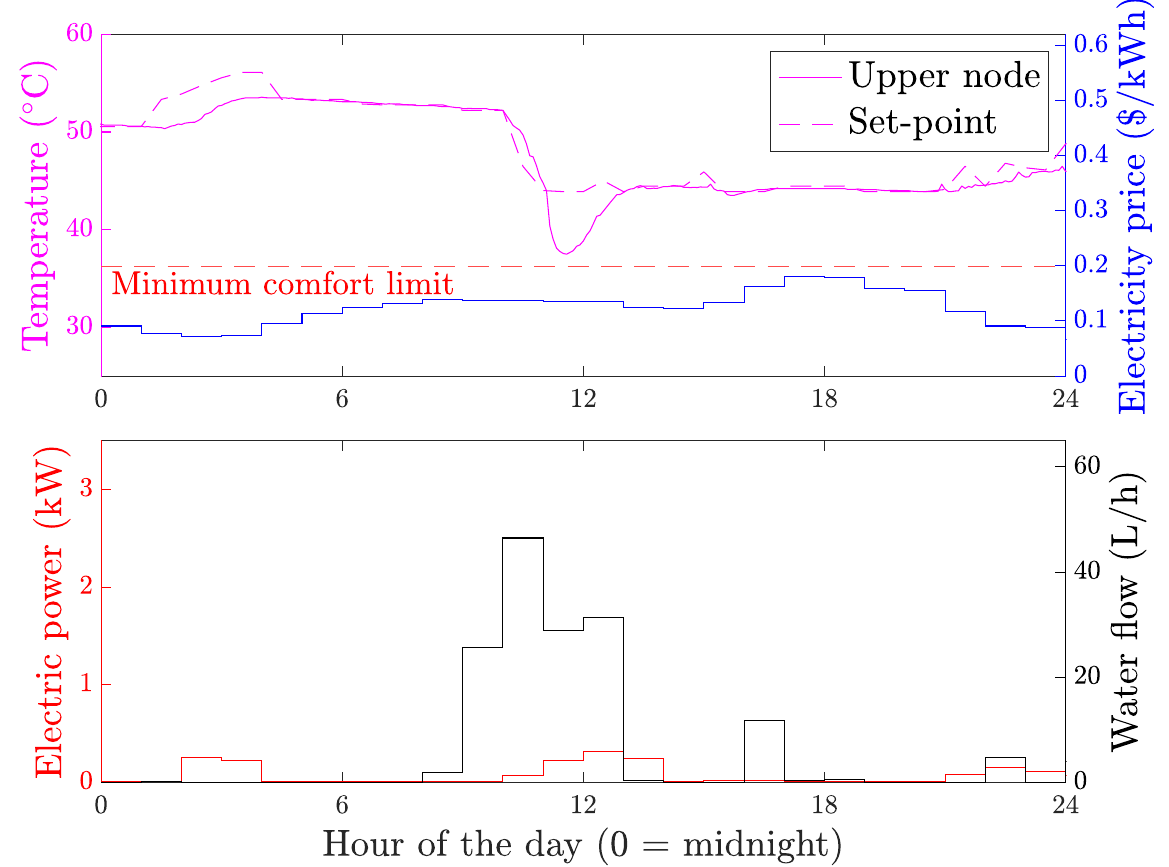}
    \caption{Under hourly pricing, the {\color{magenta} HPOWH with MPC} preheats during the lowest-price period from 4--6 AM in anticipation of morning water draws, drifts during the morning mid-peak, reheats from 11 AM to 2 PM, drifts during the evening price peak, and reheats after 9 PM when the price falls.}
    \label{fig:day_ahead_example}
\end{subfigure}
\caption{Examples of MPC operation under varying electricity pricing structures.}
\label{fig:pricing_examples}
\end{figure}

Figure \ref{fig:tou_example} shows an example day of MPC operating the HPOWH under the two-tier TOU rate. After the morning water draws, MPC raised the water temperature set-point, preheating in anticipation of the peak price period (shaded grey area). During the peak price period, there was a large water draw that would normally have triggered reheating. However, MPC dropped the set-point to avoid activating the heat pump, delaying most reheating until after the peak price period. (The heat pump was briefly activated around 4 PM due to a delay in communicating a set-point change.)

Relative to the hybrid-HPWH, MPC achieved greater cost savings with TOU rates than with a flat rate. To show this, we compared costs under TOU rates between the MPC-operated HPOWH and the hybrid-HPWH on days with similar water draws. While MPC used slightly more energy (13.7 Wh/L vs. 13.5 Wh/L), it reduced energy costs by 22.8\%, averaging 0.00142 \$/L vs. 0.00184 \$/L for the hybrid-HPWH. By shifting heating away from peak periods, MPC also offers potential benefits to the power grid, such as reducing peak demand and shifting load away from wholesale energy price spikes.

\subsection{Hourly pricing case study}
\label{hourlyPricing}

The third rate structure evaluated was hourly pricing. Prices were sourced from the local wholesale market operator, the Midcontinent Independent System Operator \cite{miso_market_displays}. In this rate structure, electricity prices vary hourly and are published once per day at 3 PM. To enable fair cost analysis, an offset was added to the wholesale prices such that the time-average price equaled the flat price from Section \ref{constantPrice}. MPC was tested with hourly prices from May 8-12, 2024.

Figure \ref{fig:day_ahead_example} shows an example day of MPC operating the HPOWH under hourly pricing. MPC preheated the water from 2--4 AM, when electricity prices were lowest. During the large water draws from 9 AM to noon, MPC reheated the water, then maintained a reduced set-point to avoid operating the heat pump during the evening price spike. MPC raised the set-point after 10 PM, reheating the water after prices dropped.

To compare costs under day-ahead hourly pricing between the MPC-operated HPOWH and the hybrid-HPWH, test days with similar water draw profiles were selected. The MPC-operated HPOWH reduced costs by 27.9\%, averaging 0.00135 \$/L vs. 0.00187 \$/L for the hybrid-HPWH. Although the MPC-operated HPOWH used slightly more energy (13.3 Wh/L vs. 13.1 Wh/L), its ability to shift load to lower-priced periods reduced costs significantly.

\section{Discussion}
\label{sec:discussion}

The results from Sections \ref{constantPrice}--\ref{hourlyPricing} show that MPC achieved greater energy cost reductions under more complex electricity pricing structures. Under a constant price, MPC increased energy costs by about 5\% relative to the hybrid-HPWH with default controls. However, as the rate structures became more dynamic, MPC’s load-shifting ability led to more pronounced energy cost savings: 22.8\% under TOU pricing and 27.8\% under hourly pricing. To complement these energy cost results, Section \ref{deploymentCost} discusses up-front costs for various HPWH configurations. Section \ref{monthlyCosts} uses the experiment data to estimate monthly energy costs. Section \ref{lifecycleEconomics} analyzes the simple payback period, based on the up-front costs and the expected annual energy cost savings. Section \ref{limitations} discusses limitations of this paper and possible directions for future work.

\subsection{Deployment cost and effort}
\label{deploymentCost}

Deploying the MPC algorithm on the HPOWH required additional sensing and computing, as detailed in Section \ref{sec:Test House}. Replicating this implementation on another water heater -- particularly as a third-party solution -- would require installing a water flow meter, an inlet temperature sensor (thermocouple or thermistor), and an edge device capable of both sensor interfacing and solving MPC optimization problems.

Water flow meters vary significantly in price, with low-cost models available around \$15 and more accurate models exceeding \$100. While hardware cost is a factor, the primary challenge with flow meter deployment lies in the labor required for installation. These devices typically require light plumbing modifications to integrate into the existing piping. Recent work suggests that temperature-based flow estimation methods can eliminate the need for a flow meter \cite{ReyesPremer2025_MinimalSensing_HPWH}, but more empirical testing is needed to demonstrate reliable comfort delivery and demand response capabilities.

The inlet water temperature can be measured via a thermocouple or thermistor, typically ranging from \$5 to \$30, which can be affixed to the inlet pipe at a metal contact point. While it is possible to install sensors in direct contact with water for higher accuracy, such installations demand additional technical expertise and labor.

For local computation and data acquisition, a Raspberry Pi or similar edge computing device can be employed. At approximately \$80 \cite{raspberrypi}, these devices provide sufficient capability to solve MPC optimization problems and store sensor data. Before deployment, hot water draw data must be collected to build a training dataset of household usage patterns. Our field demonstration used one month of data for initial model fitting, then retrained models nightly. However, initial deployment could likely proceed after approximately two weeks of data collection. Although our one-off field deployment entailed manual effort for hyperparameter tuning and model selection, we structured the algorithms such that the full machine learning pipeline could likely be automated in future deployments.

The water heater thermal model used here requires only one round of parameter tuning per equipment type. A single tuned model could be reused across all the water heaters within a product line. This eliminates the need for repeated tuning in each new deployment. Key parameters such as tank volume, thermal resistance, heat pump capacity, stratification height, COP, and heat distribution across nodes can be pre-configured. In regions with significant seasonal variation, the parameter tuning method described in Section~\ref{sec: ParameterTuning} could be applied seasonally to update the model parameters.

Manufacturers aiming to integrate MPC into their existing product lines could streamline deployment by embedding sensors and tuning processes into the production workflow. For example, flow sensors and temperature sensors could be installed in the factory rather than in the field. Parameter tuning could be completed in-house during product development. Computing resources could either be installed in the factory or shifted to cloud-based services.

\subsection{Monthly energy and cost savings}
\label{monthlyCosts}

For HPOWHs to capture a reasonable share of the commercial market, they must offer sufficient cost savings to replace hybrid-HPWHs or ERWHs, while also maintaining occupant comfort. This section evaluates the operating costs of three configurations: an MPC-operated HPOWH, a hybrid-HPWH, and a HPOWH maintained at a constant 60 $^\circ$C -- the maximum set-point of the unit. The 60 $^\circ$C HPOWH serves as a baseline representing a straightforward strategy to ensure comfort through consistently high temperatures. A HPOWH operated at 48.8 $^\circ$C was excluded due to its inability to reliably maintain comfort, making it a poor candidate for market deployment.

\begin{figure}
\centering
\includegraphics[width=0.5\linewidth]{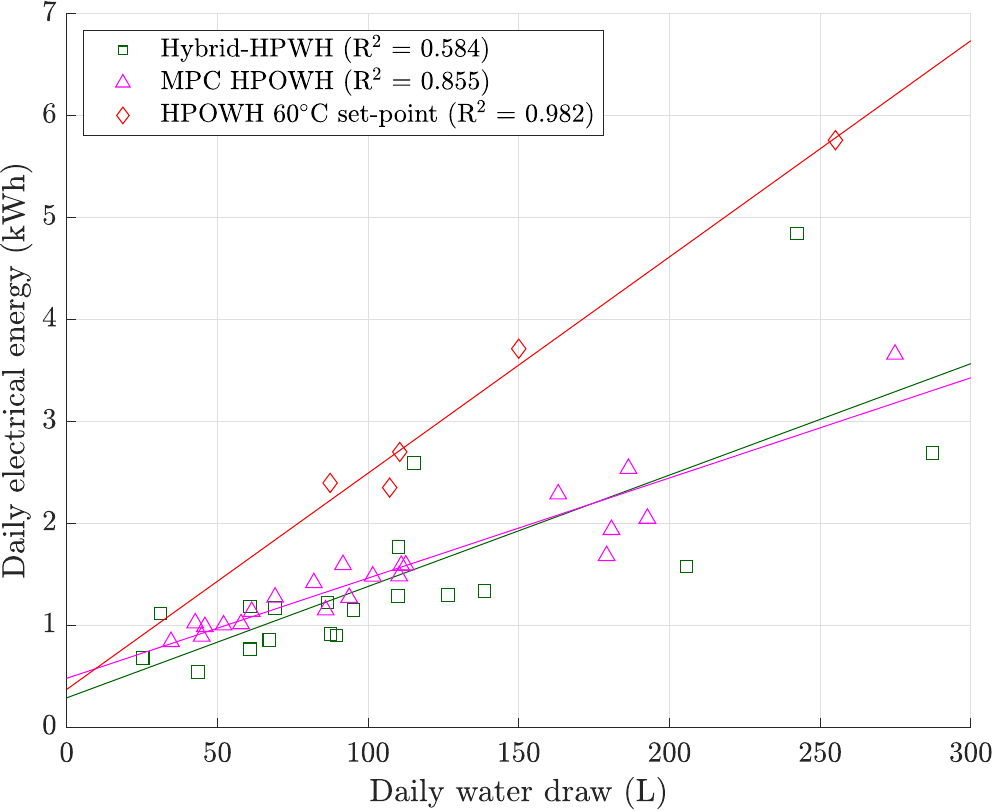}
\caption{The {\color{magenta} HPOWH with MPC} uses much less energy for a given daily water draw than the {\color{red} HPOWH at 60$^{\circ}$C}. The {\color{ForestGreen} hybrid-HPWH with default controls} uses similar energy to the {\color{magenta} HPOWH with MPC} on most days, but occasionally uses much more energy due to its inefficient heating elements.}
\label{fig:linearRegres}
\end{figure}

Linear regression models were applied to experimental data to estimate summer operating costs for each water heater mode. Figure \ref{fig:linearRegres} shows the data points and linear regression fits. To evaluate the fits, the coefficient of determination (R$^2$) -- a common measure of goodness of fit for linear regression models -- was calculated. The MPC-operated HPOWH and the HPOWH held at a constant 60 $^\circ$C show clear linear trends with high R$^2$ values. The HPOWH held at 60 $^\circ$C uses almost twice as much energy as the MPC HPOWH because increasing the water temperature decreases the heat pump COP and increases conductive heat losses through the tank wall. The hybrid-HPWH, while still exhibiting a general linear trend, had a lower R$^2$ value of 0.584, largely due to sporadic activation of its resistance heating elements. Despite this variability, the regression models are suitable for extrapolating summer performance from observed data.

Data from May through September 2024 were used to project summer energy costs. Because the home experiences cold winters, only summer data -- when inlet temperatures are more stable -- were considered reliable for extrapolation. The effects of colder winter inlet conditions should be investigated in future work. Periods with missing data exceeding one week were filled using randomized samples from similar fully occupied days. The average monthly water draw was 3,710 L, ranging over months from 2,730 to 5,040 L.

\begin{table*}
\small
\centering
\caption{Monthly energy and cost comparison for HPWH operating modes over the summer (May--September)}
\label{table:cost_comparison}
\begin{tabular}{lcc}
& \textbf{MPC vs. 60$^{\circ}$C set-point} & \textbf{MPC vs. Hybrid} \\
\midrule
Mean monthly energy savings (and range) & 37.8 (22.1--54.1) kWh & -1.89 (0.4--3.5) kWh \\
Mean monthly percent energy savings (and range)  & 37.6 (28.6-46.6)\%               & -3.9 (0.6-7.6)\% \\
\midrule
Mean monthly cost savings at 0.13 \$/kWh & \$3.78 (\$2.24--5.40) & -\$0.19 (\$0.04--0.35) \\
Mean monthly cost savings at 0.20 \$/kWh & \$7.57 (\$4.48--10.81) & -\$0.38 (\$0.08--0.69) \\
Mean monthly cost savings at 0.30 \$/kWh & \$11.36 (\$6.72--16.21) & -\$0.57 (\$0.12--1.04) \\
\end{tabular}
\end{table*}

Table \ref{table:cost_comparison} shows energy usage and cost comparisons derived from the linear regression models. Comparing the MPC-operated HPOWH to the constant 60 $^\circ$C HPOWH, the relative energy savings average 37.6\% during the summer months, with mean monthly energy savings of 37.8 kWh. For the three-person household, the estimated monthly cost savings are relatively small with a low flat rate of electricity, but  reach an average of \$11.36 (\$6.72–\$16.21) per month at a higher rate of 0.30 \$/kWh \cite{poweroutage2025}.

When comparing the MPC-operated HPOWH to the hybrid-HPWH, MPC used slightly more energy due to higher average water temperatures. The relative increase was 3.9\%, corresponding to an additional 1.89 kWh per month on average and a monthly energy cost difference of less than \$1. Although these energy cost differences are negligible, the installation cost differences between the two systems are potentially significant, as discussed in the next section.

\subsection{Simple payback analysis}
\label{lifecycleEconomics}

Replacing a gas water heater with a hybrid-HPWH typically requires installing a dedicated 240 V circuit due to the resistance heating elements' high power draws. Equipment costs for hybrid-HPWHs average \$1,700 \cite{NBI2023}, with installation costs commonly adding around \$3,000, although this can vary significantly from \$2,000 to \$4,600 based on local labor and market conditions \cite{NBI2023}. If an electrical panel upgrade is also needed, due either to insufficient total current capacity or to a lack of available 240 V circuit breaker slots, installation costs increase by an average of \$3,000 (ranging from \$2,000 to \$5,000) \cite{rewiringamerica2024}. Therefore, total up-front costs -- including equipment, installation, and any required electrical work -- range approximately from \$4,700 (with an existing 240 V circuit and no panel upgrade), to around \$5,900 (without an existing 240 V outlet or panel upgrade), and \$8,500 or higher if a panel upgrade is required.

The equipment cost of a 120 V shared-circuit HPOWH is typically around \$2,300. The \$600 average equipment cost increase over a hybrid-HPWH comes mainly from HPOWHs' use of higher-capacity heat pumps and of mixing valves for scald protection \cite{NBI2023}.  Installation for 120 V HPOWHs averages around \$3,000, with a possible additional \$100 if an outlet needs to be extended. Total up-front costs are therefore typically \$5,300 to \$5,400.

Compared to the typical up-front cost of a hybrid-HPWH that requires a dedicated 240 V outlet (around \$5,900), the 120 V shared-circuit HPOWH offers about \$500 to \$600 in up-front cost savings. If a panel upgrade is avoided, savings increase to \$3,200 or more. As the energy cost differences between a hybrid-HPWH and an MPC-operated HPOWH are negligible, the up-front cost savings make an MPC-operated HPOWH the economically preferable option as long as MPC adds no more than \$500 to the up-front cost. Based on the discussion in Section \ref{deploymentCost}, we expect MPC deployment costs to be well below \$500. Beyond up-front cost savings relative to hybrid-HPWHs, an MPC-operated HPOWH also offers load-shifting capabilities that could earn revenue through participation in utility demand response programs. Demand response revenues are not included here. For homeowners planning further electrification (e.g., stoves or EV charging), leaving more room in the panel for other loads and avoiding future panel upgrades also provides value not included here.

To compare an MPC-operated HPOWH to a HPOWH that stores water at 60 $^\circ$C, we assume that any additional sensing and computing required for MPC costs at most \$200 per water heater. Under this assumption, the MPC-operated HPOWH costs \$200 more up front, but saves money over time on energy bills. Based on the mean monthly energy cost savings in Table \ref{table:cost_comparison}, the estimated simple payback period for an MPC-operated HPOWH relative to a HPOWH that stores water at 60 $^\circ$C is approximately 53 months with an electricity price of 0.13 \$/kWh, 26 months at 0.20 \$/kWh, and 18 months at 0.30 \$/kWh. Under more dynamic rate structures -- such as time-of-use or real-time pricing -- MPC is expected to yield greater savings by strategically shifting energy consumption to lower-cost periods. MPC could also unlock demand response revenues that are not included here.

In summary, the calculations in this section suggest that an MPC-operated HPOWH is economically preferable to both a hybrid-HPWH, due to up-front cost savings, and to a HPOWH that stores water at 60 $^\circ$C, due to energy cost savings. The calculations in this section do not include the additional economic benefits that MPC could unlock from arbitraging time-varying energy prices or participating in utility demand response programs.

\subsection{Limitations and future work}
\label{limitations}

Deployment at scale of the cyber-physical system developed in this paper would require extending the work in at least three directions. First, the reliability and cost of the IoT system should be improved. Initially, robust communication protocols and automatic recovery from system failures were not implemented, leading to occasional data losses during field testing due to IoT disruptions. These disruptions arose from various factors, including edge device software bugs, power outages, and WiFi network resets. While specialized expertise can mitigate these challenges, such skills may not be commonly found within existing water heater manufacturers. Implementing MPC via a third-party solution, as done in this field test, introduces additional complexity, such as integration challenges with existing manufacturer controls, supplementary sensor installations, and ensuring reliable API connectivity with different brands of water heaters.

Second, the water flow meter could be eliminated. Although it was valuable for accurate measurements in this research study, the flow meter's cost and installation challenges may be prohibitive at scale. Future work could explore replacing the physical flow meter with a flow estimation algorithm that is sufficiently accurate to enable hot water draw forecasting and predictive control. Recent work suggests that temperature-based flow estimation is a practical path forward \cite{Buechler_EWH2_2024, ReyesPremer2025_MinimalSensing_HPWH, delaRosa_estimation}, but this should be verified in the field.

Third, MPC caused more frequent compressor cycling due to its dynamic set-point adjustments, exacerbated by communication delays and model inaccuracies. Additional cycling increases wear and tear on the compressor and could shorten its life. While the linear optimization approach used here has benefits for computational efficiency and scalability, it likely prohibits directly including cycling objectives in the MPC optimization. Future research could  compare the linear formulation to mixed-integer formulations that explicitly penalize cycling \cite{DELAROSA202583}, assessing trade-offs between cycling and computational complexity.

\section{Conclusion}
\label{sec:conclusion}

Water heaters powered by electricity are important for residential decarbonization. Among the available technologies, heat-pump water heaters are the most efficient option and are likely to comprise the main technology pathway in the United States. However, the high up-front costs associated with 240 V hybrid-heat-pump water heaters pose a significant barrier to widespread adoption. Heat-pump-only water heaters offer lower up-front costs, as they require only a standard 120 V outlet.

To date, concerns related to occupant comfort and energy efficiency have limited the appeal of 120 V heat-pump-only water heaters. To address these challenges, this paper developed a model predictive control system that maintained comfortable water temperatures while reducing energy costs. The control system adapted well to various electricity pricing structures, including flat rates, time-of-use rates, and hourly pricing. Economic analysis, informed by field data, suggested that the control system is the most attractive option in several common deployment settings. \textcolor{black}{Although this study focused on a water heater in a single home, the modeling, forecasting, and control framework was designed for broader applicability. Future work should validate the approach on different water heaters and in varied climate zones, quantifying deployment effort, achievable energy and cost savings, and comfort performance.}

\section*{CRediT authorship contribution statement}

{\bf Levi D. Reyes Premer:} Conceptualization, Methodology, Investigation, Formal Analysis, Data Curation, Visualization, Writing – Original Draft, Writing – Review \& Editing. {\bf Elias N. Pergantis:} Software, Methodology, Formal Analysis, Data Curation, Visualization, Writing – Review \& Editing. {\bf Leo Semmelmann:} Conceptualization, Software, Methodology, Visualization, Writing – Original Draft, Writing – Review \& Editing. {\bf Davide Zivani:} Writing - Review \& Editing, Project Administration, Funding Acquisition. {\bf Kevin Kircher:} Conceptualization, Methodology, Writing - Review \& Editing, Project Administration, Funding Acquisition.

\section*{Nomenclature}

{\scriptsize
\setlength{\tabcolsep}{6pt}
\renewcommand{\arraystretch}{1.1}

\noindent
\begin{minipage}[t]{0.49\linewidth}
\begin{tabularx}{\linewidth}{@{}l|Y@{}}
\textbf{Symbol} & \textbf{Description} \\ \hline
$A$               & Discrete-time system matrix \\
$\tilde{A}$       & Continuous-time system matrix \\
$A_{\text{WH}}$   & Tank cross-sectional area (m$^2$) \\
$a$               & Set-point tracking parameter \\
$B$               & Discrete-time input matrix \\
$\tilde{B}$       & Continuous-time input matrix \\
$c_p$             & Specific heat of water (kWh/(kg$\cdot^\circ$C)) \\
$c_{\text{elec}}$ & Electricity price (\$/kWh) \\
$C$               & Tank thermal capacitance (kWh/$^{\circ}$C) \\
$h$               & Total tank height (m) \\
$h_{\text{thrm}}$ & Thermocline height (m) \\
$h_s$             & Stratification layer thickness (m) \\
$I$               & Identity matrix \\
$J$               & Prediction horizon length (time-steps) \\
$J_1, J_2$        & Horizon thresholds for ensemble model \\
$k$               & Discrete-time index \\
$k_w$             & Thermal conductivity of water (W/m/$^{\circ}$C) \\
$\lambda$         & Fraction of heat output to upper node \\
$\dot{m}$         & Mass flow rate (kg/h) \\
$\hat{\dot{m}}$   & Forecasted mass flow rate (kg/min) \\
$P$               & Electrical power input to heat pump (kW) \\
$P_{\max}$        & Maximum electrical power input (kW) \\
\end{tabularx}
\end{minipage}\hfill
\begin{minipage}[t]{0.49\linewidth}
\begin{tabularx}{\linewidth}{@{}l|Y@{}}
\textbf{Symbol} & \textbf{Description} \\ \hline
$q$               & Heat pump thermal output (kW) \\
$R_a$             & Tank/ambient thermal resistance ($^{\circ}$C/kW) \\
$R_{u\ell}$       & Upper/lower thermal resistance ($^{\circ}$C/kW) \\
$T_a$             & Ambient air temperature ($^{\circ}$C) \\
$\hat{T}_a$       & Forecasted ambient temperature ($^{\circ}$C) \\
$T_c$             & Inlet tank water temperature ($^{\circ}$C) \\
$\hat{T}_c$       & Forecasted inlet water temperature ($^{\circ}$C) \\
$T_{\ell}$        & Lower-node temperature ($^{\circ}$C) \\
$T_u$             & Upper-node temperature ($^{\circ}$C) \\
$T_{\min}$        & Minimum water temperature ($^{\circ}$C) \\
$T_s$             & Temperature set-point ($^{\circ}$C) \\
$T_{s,\max}$      & Maximum set-point temperature ($^{\circ}$C) \\
$T_{\text{bact}}$ & Bacteria growth threshold temperature ($^{\circ}$C) \\
$\Delta t$        & Time step duration \\
$\eta$            & Coefficient of performance (COP) \\
$\gamma$          & Comfort/bacteria penalty weight (\$/$^{\circ}$C/h) \\
$\pi$             & Indicator for bacteria growth penalty \\
$\varphi$         & Flow threshold for large water draw (kg/min) \\
$\Phi$            & Rolling draw volume threshold (kg) \\
$w$               & Discrete-time disturbance vector \\
$\tilde{w}$       & Continuous-time disturbance vector \\
$z$               & Fraction of tank height in upper node \\
\end{tabularx}
\end{minipage}
}

\section*{Declaration of competing interest}

The authors declare that they have no known competing financial interests or personal relationships that could have appeared to influence the work reported in this paper.

\section*{Data availability}

Data will be made available on request.

\section*{Acknowledgments}

The Center for High-Performance Buildings at Purdue University supported this work. Levi D. Reyes Premer was also supported by the NSF Graduate Research Fellowship. Elias Pergantis was supported by an Onassis Foundation scholarship. Levi and Elias were both supported by the American Society of Heating and Refrigeration Engineers (ASHRAE) through the Grant-In-Aid award. The authors would like to thank the occupants of the test house for their patience and cooperation during testing.

\bibliographystyle{elsarticle-num} 
\bibliography{cas-refs}

\begin{thebibliography}{10}
\expandafter\ifx\csname url\endcsname\relax
  \def\url#1{\texttt{#1}}\fi
\expandafter\ifx\csname urlprefix\endcsname\relax\def\urlprefix{URL }\fi
\expandafter\ifx\csname href\endcsname\relax
  \def\href#1#2{#2} \def\path#1{#1}\fi

\bibitem{CEC2024EnergyCode}
{California Energy Commission}, Energy commission adopts updated building
  standards expanding requirements for heat pumps and electric-ready buildings
  (2024).

\bibitem{DOE2029HPWH}
{DOE}, {DOE} finalizes efficiency standards for water heaters to save americans
  over \$7 billion on household utility bills annually (2024).

\bibitem{RECS2020}
EIA, Residential energy consumption survey ({RECS}) (2020).

\bibitem{NRELHPWH2012}
K.~Hudson, B.~Sparn, D.~Christensen, J.~Maguire, Heat pump water heater
  technology assessment based on laboratory research and energy simulation
  models, Tech. rep., {NREL} (2012).

\bibitem{maguire_comparison_2013}
J.~Maguire, X.~Fang, E.~Wilson, Comparison of advanced residential water
  heating technologies in the {United States}, {Renewable Energy} (2013).

\bibitem{priyadarshan2024edgie}
Priyadarshan, E.~N. Pergantis, C.~Crozier, K.~Baker, K.~J. Kircher, {EDGIE}: A
  simulation test-bed for investigating the impacts of building and vehicle
  electrification on distribution grids, {Proceedings of the Hawaii
  International Conference on System Sciences} (2024).

\bibitem{BILLERBECK2024117850}
A.~Billerbeck, C.~P. Kiefer, J.~Winkler, C.~Bernath, F.~Sensfuß, L.~Kranzl,
  A.~Müller, M.~Ragwitz, The race between hydrogen and heat pumps for space
  and water heating: {A} model-based scenario analysis, {Energy Conversion and
  Management} 299 (2024) 117850.

\bibitem{ENERGYSTAR2022_HPWHSales}
{Energy Star}, {Energy Star®} unit shipment and market penetration report,
  calendar year 2022 summary (2022).

\bibitem{ENERGYSTAR2024_HPWH}
{Energy Star}, Heat pump water heater market acceleration guide (2024).

\bibitem{REWIREAMRERICA_HPWH}
{Rewiring America}, Report: Upfront cost of home electrification (2024).

\bibitem{SATREMELOY2024110939}
A.~Satre-Meloy, N.~Casquero-Modrego, B.~Less, I.~Walker, Reducing the cost of
  home energy upgrades in the {US}: An industry survey, {Journal of Building
  Engineering} 98 (2024) 110939.

\bibitem{pergantis_current}
E.~N. Pergantis, L.~D.~R. Premer, A.~H. Lee, Priyadarshan, H.~Liu, E.~Groll,
  D.~Ziviani, K.~J. Kircher, Protecting residential electrical infrastructure
  through advanced control: The first field results, {8th International High
  Performance Buildings Conference at Purdue} (2024).

\bibitem{pergantis_current_APEN}
E.~N. Pergantis, L.~D.~R. Premer, A.~H. Lee, Priyadarshan, H.~Liu, E.~A. Groll,
  D.~Ziviani, K.~J. Kircher, Protecting residential electrical panels and
  service through model predictive control: A field study, {Applied Energy}
  (2024).

\bibitem{osti_885625}
B.~Ashdown, Heat pump water heater technology: Experiences of residential
  consumers and utilities, Tech. rep., {Oak Ridge National Laboratory} (2004).

\bibitem{efficiencyfirstca_heat_pump_water_heaters}
{Efficiency First California}, Heat pump water heaters - we need to get this
  right! (2023).

\bibitem{mande2022timing}
C.~Mande, A.~Aboud, L.~dela Rosa, et~al., Timing is everything: Optimizing load
  flexibility of heat pump water heaters for cost, comfort, and carbon
  emissions, {UC Davis} (2022).

\bibitem{shen2021data}
G.~Shen, Z.~E. Lee, A.~Amadeh, K.~M. Zhang, A data-driven electric water heater
  scheduling and control system, {Energy and Buildings} 242 (2021) 110924.

\bibitem{dela2021supervisory}
L.~dela Rosa, C.~Mande, M.~J. Ellis, Supervisory multi-objective economic model
  predictive control for heat pump water heaters for cost and carbon
  optimization, {ASHRAE Winter Conference} (2023).

\bibitem{BAUMANN2023112923}
C.~Baumann, G.~Huber, J.~Alavanja, M.~Preißinger, P.~Kepplinger, Experimental
  validation of a state-of-the-art model predictive control approach for demand
  side management with a hot water heat pump, {Energy and Buildings} 285 (2023)
  112923.

\bibitem{Jin2014}
X.~Jin, J.~Maguire, D.~Christensen, Model predictive control of heat pump water
  heaters for energy efficiency (08 2014).

\bibitem{Bastian2022}
H.~Bastian, C.~Cohn, Ready to upgrade: Barriers and strategies for residential
  electrification, {American Council for an Energy-Efficient Economy (ACEEE)}
  (2022).

\bibitem{TARROJA2018522}
B.~Tarroja, F.~Chiang, A.~AghaKouchak, S.~Samuelsen, S.~V. Raghavan, M.~Wei,
  K.~Sun, T.~Hong, Translating climate change and heating system
  electrification impacts on building energy use to future greenhouse gas
  emissions and electric grid capacity requirements in california, {Applied
  Energy} 225 (2018) 522--534.

\bibitem{Decarb2050DOE}
{DOE}, Decarbonizing the u.s. economy by 2050 (2024).

\bibitem{ZHANG2019709}
L.~Zhang, N.~Good, P.~Mancarella, Building-to-grid flexibility: Modelling and
  assessment metrics for residential demand response from heat pump
  aggregations, {Applied Energy} 233-234 (2019) 709--723.

\bibitem{LACROIX19991313}
M.~Lacroix, Electric water heater designs for load shifting and control of
  bacterial contamination, {Energy Conversion and Management} 40~(12) (1999)
  1313--1340.

\bibitem{DELAROSA202583}
L.~{dela Rosa}, C.~Mande, M.~J. Ellis, Beyond the one-shift wonder: A case
  study on predictive control for heat pump water heaters, {Chemical
  Engineering Research and Design} 215 (2025) 83--97.

\bibitem{EARLE2023120256}
L.~Earle, J.~Maguire, P.~Munankarmi, D.~Roberts, The impact of
  energy-efficiency upgrades and other distributed energy resources on a
  residential neighborhood-scale electrification retrofit, {Applied Energy} 329
  (2023) 120256.

\bibitem{ZHAO2024119026}
Z.~Zhao, B.~Wang, X.~Li, W.~Shi, Adaptive model predictive control of a
  residential solar-air hybrid heat pump system, {Energy Conversion and
  Management} 321 (2024) 119026.

\bibitem{YANG2021114710}
L.~W. Yang, R.~J. Xu, N.~Hua, Y.~Xia, W.~B. Zhou, T.~Yang, Y.~Belyayev, H.~S.
  Wang, Review of the advances in solar-assisted air source heat pumps for the
  domestic sector, {Energy Conversion and Management} 247 (2021) 114710.

\bibitem{DISILVESTRE2023113425}
M.~{Di Silvestre}, E.~{Riva Sanseverino}, E.~Telaretti, G.~Zizzo, Flexibility
  of grid interactive water heaters: The situation in the {US}, {Renewable and
  Sustainable Energy Reviews} 182 (2023) 113425.

\bibitem{cta2045report2018}
B.~P. Administration, Demand response market transformation and business case
  report (2018).

\bibitem{CTA2045B2022}
{Consumer Technology Association (CTA)}, {ANSI/CTA-2045-B}: Modular
  communication interface for energy management (2022).

\bibitem{Jordan2001}
U.~Jordan, K.~Vajen, Influence of the dhw load profile on the fractional energy
  savings, {Solar Energy} 69 (2001) 197--208.

\bibitem{RITCHIE2021110727}
M.~Ritchie, J.~Engelbrecht, M.~Booysen, A probabilistic hot water usage model
  and simulator for use in residential energy management, {Energy and
  Buildings} 235 (2021) 110727.

\bibitem{EDWARDS201543}
S.~Edwards, I.~Beausoleil-Morrison, A.~Laperrière, Representative hot water
  draw profiles at high temporal resolution for simulating the performance of
  solar thermal systems, {Solar Energy} 111 (2015) 43--52.

\bibitem{Cao2019}
S.~Cao, S.~Hou, L.~Yu, J.~Lu, Predictive control based on occupant behavior
  prediction for domestic hot water system using data mining algorithm, {Energy
  Science \& Engineering} 7~(4) (2019) 1214--1232.

\bibitem{PFLUGRADT2017655}
N.~Pflugradt, U.~Muntwyler, Synthesizing residential load profiles using
  behavior simulation, {Energy Procedia} 122 (2017) 655--660.

\bibitem{Gelazanskas2015}
L.~Gelazanskas, K.~Gamage, Forecasting hotwater consumption in residential
  houses, Energies 8 (2015) 12702--12717.

\bibitem{CLIFT2023126577}
D.~H. Clift, C.~Stanley, K.~N. Hasan, G.~Rosengarten, Assessment of advanced
  demand response value streams for water heaters in renewable-rich electricity
  markets, Energy 267 (2023) 126577.

\bibitem{RITCHIE2024123421}
M.~Ritchie, J.~Engelbrecht, M.~Booysen, Loadshedding-induced transients due to
  battery backup systems and electric water heaters, {Applied Energy} 367
  (2024) 123421.

\bibitem{Buechler_Goldin_EWH1_2024}
E.~Buechler, A.~Goldin, R.~Rajagopal, Improving the load flexibility of
  stratified electric water heaters: Design and experimental validation of mpc
  strategies, {IEEE Transactions on Smart Grid} 15~(4) (2024) 3613–3623.

\bibitem{2nodeHmodel_DIAO1012}
R.~Diao, S.~Lu, M.~Elizondo, E.~Mayhorn, Y.~Zhang, N.~Samaan, Electric water
  heater modeling and control strategies for demand response, {Power and Energy
  Society General Meeting, 2012 IEEE} (2012) 1--8.

\bibitem{LUO2024110724}
Z.~Luo, J.~Peng, X.~Zhang, H.~Jiang, M.~Lv, Load flexibility quantification of
  electric water heaters under various demand-side management strategies and
  seasons, {Journal of Building Engineering} 97 (2024) 110724.

\bibitem{Lin_Li_Xiao_2017}
B.~Lin, S.~Li, Y.~Xiao, Optimal and learning-based demand response mechanism
  for electric water heater system, Energies 10~(1111) (2017) 1722.

\bibitem{ELBAKALI2024118190}
S.~{El Bakali}, H.~Ouadi, S.~Gheouany, Efficient real-time cost optimization of
  a two-layer electric water heater system under model uncertainties, {Energy
  Conversion and Management} 304 (2024) 118190.

\bibitem{GAONWE2022}
T.~P. Gaonwe, K.~Kusakana, P.~A. Hohne, A review of solar and air-source
  renewable water heating systems, under the energy management scheme, {Energy
  Reports} 8 (2022) 1--10.

\bibitem{DELAROSA2025_Rule}
L.~{dela Rosa}, C.~Mande, M.~J. Ellis, Practical strategies for managing
  resistance heating in heat pump water heater predictive control, {Chemical
  Engineering Research and Design} 215 (2025) 180--192.

\bibitem{en12030411}
A.~Oshnoei, R.~Khezri, S.~M. Muyeen, Model predictive-based secondary frequency
  control considering heat pump water heaters, Energies 12~(3) (2019).

\bibitem{amasyali_deep_2021}
K.~Amasyali, J.~Munk, K.~Kurte, T.~Kuruganti, H.~Zandi, Deep reinforcement
  learning for autonomous water heater control, Buildings 11~(11) (2021) 548.

\bibitem{yin_data-driven_2024}
M.~Yin, H.~Cai, A.~Gattiglio, F.~Khayatian, R.~S. Smith, P.~Heer, Data-driven
  predictive control for demand side management: Theoretical and experimental
  results, {Applied Energy} 353 (2024).

\bibitem{Starke2020}
M.~Starke, J.~Munk, H.~Zandi, T.~Kuruganti, H.~Buckberry, J.~Hall,
  J.~Leverette, Real-time mpc for residential building water heater systems to
  support the electric grid, {2020 IEEE Power \& Energy Society Innovative
  Smart Grid Technologies Conference (ISGT)} (2020) 1--5.

\bibitem{StarkeNeighbor2019}
M.~Starke, J.~Munk, H.~Zandi, T.~Kuruganti, H.~Buckberry, J.~Hall,
  J.~Leverette, Agent-based system for transactive control of smart residential
  neighborhoods, in: 2019 IEEE Power \& Energy Society General Meeting
  ({PESGM}), 2019, pp. 1--5.

\bibitem{Nash_Badithela_Jain_2017}
A.~L. Nash, A.~Badithela, N.~Jain, Dynamic modeling of a sensible thermal
  energy storage tank with an immersed coil heat exchanger under three
  operation modes, {Applied Energy} 195 (2017) 877–889.

\bibitem{Xu2024}
Z.~Xu, R.~Diao, S.~Lu, J.~Lian, Y.~Zhang, Modeling of electric water heaters
  for demand response: A baseline {PDE} model, {IEEE Transactions on Smart
  Grid} 5~(5) (2014) 2203--2210.

\bibitem{Buechler_EWH2_2024}
E.~Buechler, A.~Goldin, R.~Rajagopal, Designing model predictive control
  strategies for grid-interactive water heaters for load shifting applications
  (2024).

\bibitem{Bartolucci2019}
L.~Bartolucci, S.~Cordiner, V.~Mulone, M.~Santarelli, Hybrid renewable energy
  systems: Influence of short term forecasting on model predictive control
  performance, Energy 172 (2019) 997--1004.

\bibitem{LydenTuohy2022}
A.~Lyden, P.~G. Tuohy, Planning level sizing of heat pumps and hot water tanks
  incorporating model predictive control and future electricity tariffs, Energy
  238 (2022) 121731.

\bibitem{semmelmann2022load}
L.~Semmelmann, S.~Henni, C.~Weinhardt, Load forecasting for energy communities:
  A novel {LSTM-XGBoost} hybrid model based on smart meter data, {Energy
  Informatics} 5~(Suppl 1) (2022) 24.

\bibitem{winter2002shapley}
E.~Winter, The shapley value, {Handbook of game theory with economic
  applications} 3 (2002) 2025--2054.

\bibitem{ghilardi2023benefits}
A.~Ghilardi, G.~F. Frate, M.~Tucci, M.~Bravi, R.~Leo, L.~Ferrari, Benefits of
  thermal load forecasts in balancing load fluctuations through thermal
  storage, {Journal of Energy Storage} 70 (2023) 107929.

\bibitem{lazos2015development}
D.~Lazos, A.~B. Sproul, M.~Kay, Development of hybrid numerical and statistical
  short term horizon weather prediction models for building energy management
  optimisation, {Building and Environment} 90 (2015) 82--95.

\bibitem{saber2017short}
A.~Y. Saber, A.~R. Alam, Short term load forecasting using multiple linear
  regression for big data, in: {2017 IEEE symposium series on computational
  intelligence (SSCI)}, {IEEE}, 2017, pp. 1--6.

\bibitem{xu2021automl}
Z.~Xu, W.-W. Tu, I.~Guyon, Automl meets time series regression design and
  analysis of the autoseries challenge, in: {Machine Learning and Knowledge
  Discovery in Databases. Applied Data Science Track: European Conference, ECML
  PKDD 2021, Bilbao, Spain, September 13--17, 2021, Proceedings, Part V 21},
  Springer, 2021, pp. 36--51.

\bibitem{breiman2001random}
L.~Breiman, Random forests, Machine learning 45 (2001) 5--32.

\bibitem{grinsztajn2022tree}
L.~Grinsztajn, E.~Oyallon, G.~Varoquaux, Why do tree-based models still
  outperform deep learning on typical tabular data?, {Advances in neural
  information processing systems} 35 (2022) 507--520.

\bibitem{chen2016xgboost}
T.~Chen, C.~Guestrin, Xgboost: A scalable tree boosting system, in:
  {Proceedings of the 22nd acm sigkdd international conference on knowledge
  discovery and data mining}, 2016, pp. 785--794.

\bibitem{abbasi2019short}
R.~A. Abbasi, N.~Javaid, M.~N.~J. Ghuman, Z.~A. Khan, S.~Ur~Rehman, Amanullah,
  Short term load forecasting using {XGBoost}, in: {Web, Artificial
  Intelligence and Network Applications: Proceedings of the Workshops of the
  33rd International Conference on Advanced Information Networking and
  Applications (WAINA-2019) 33}, Springer, 2019, pp. 1120--1131.

\bibitem{zhang2021time}
L.~Zhang, W.~Bian, W.~Qu, L.~Tuo, Y.~Wang, Time series forecast of sales volume
  based on {XGBoost}, in: {Journal of Physics: Conference Series}, Vol. 1873,
  {IOP Publishing}, 2021, p. 012067.

\bibitem{taylor2018forecasting}
S.~J. Taylor, B.~Letham, Forecasting at scale, {The American Statistician}
  72~(1) (2018) 37--45.

\bibitem{zhou2021informer}
H.~Zhou, S.~Zhang, J.~Peng, S.~Zhang, J.~Li, H.~Xiong, W.~Zhang, Informer:
  Beyond efficient transformer for long sequence time-series forecasting, in:
  {Proceedings of the AAAI conference on artificial intelligence}, Vol.~35,
  2021, pp. 11106--11115.

\bibitem{amin2019automating}
P.~Amin, L.~Cherkasova, R.~Aitken, V.~Kache, Automating energy demand modeling
  and forecasting using smart meter data, in: {2019 IEEE International Congress
  on Internet of Things (ICIOT)}, {IEEE}, 2019, pp. 133--137.

\bibitem{vom2020data}
F.~vom Scheidt, H.~Medinov{\'a}, N.~Ludwig, B.~Richter, P.~Staudt,
  C.~Weinhardt, Data analytics in the electricity sector--a quantitative and
  qualitative literature review, {Energy and AI} 1 (2020) 100009.

\bibitem{cleger2012use}
S.~Cleger-Tamayo, J.~M. Fern{\'a}ndez-Luna, J.~F. Huete, On the use of weighted
  mean absolute error in recommender systems., in: RUE@ RecSys, 2012, pp.
  24--26.

\bibitem{showerTemp2}
C.~Herrmann, V.~Candas, A.~Hoeft, I.~Garreaud, Humans under showers: Thermal
  sensitivity, thermoneutral sensations, and comfort estimates, {Physiology \&
  Behavior} 56~(5) (1994) 1003--1008.

\bibitem{Legionella_Temp}
{National Academies of Sciences, Engineering, and Medicine}, Management of
  Legionella in Water Systems: Strategies for Legionella Control and Their
  Application in Building Water Systems, {National Academies Press (US)},
  Washington, DC, 2019.

\bibitem{MATLAB}
{The MathWorks Inc.}, \href{https://www.mathworks.com}{{MATLAB} version: 9.13.0
  (r2022b)} (2022).
\newline\urlprefix\url{https://www.mathworks.com}

\bibitem{gurobi}
{Gurobi Optimization, LLC}, \href{https://www.gurobi.com}{{Gurobi Optimizer
  Reference Manual}} (2024).
\newline\urlprefix\url{https://www.gurobi.com}

\bibitem{cvx}
M.~Grant, S.~Boyd, {CVX}: Matlab software for disciplined convex programming,
  version 2.1, \url{https://cvxr.com/cvx} (mar 2014).

\bibitem{CHPB_DCNanogridHouse}
{Center for High Performance Buildings, Purdue University}, {DC Nanogrid
  House}, \url{https://chpb.engineering.purdue.edu/dc-house-nanogrid/},
  accessed: 2025-08-25 (2025).

\bibitem{RheemHPWH}
Rheem,
  \href{https://www.rheem.com/products/residential/water-heating/heat-pump-water-heaters/}{Heat
  pump water heaters} (2024).
\newline\urlprefix\url{https://www.rheem.com/products/residential/water-heating/heat-pump-water-heaters/}

\bibitem{uefrating}
{Energy Star}, Water heater key product criteria (2023).

\bibitem{influxdb}
{InfluxDB} 2.0, {Accessed}: 03/02/2024 (2024).

\bibitem{pergantis2024field}
E.~N. Pergantis, Priyadarshan, N.~A. Theeb, P.~Dhillon, J.~P. Ore, D.~Ziviani,
  E.~A. Groll, K.~J. Kircher, Field demonstration of predictive heating control
  for an all-electric house in a cold climate, {Applied Energy} 360 (2024)
  122820.

\bibitem{LUOComfort2023}
M.~Luo, S.~Xu, Y.~Tang, H.~Yu, X.~Zhou, Z.~Chen, Dynamic thermal responses and
  showering thermal comfort under different conditions, {Building and
  Environment} 237 (2023).

\bibitem{tipmont2025tou}
{Tipmont REMC}, Time-of-use rates (2024).

\bibitem{miso_market_displays}
{Midcontinent Independent System Operator, Inc. (MISO)}, {MISO} market displays
  (day‑ahead and real‑time market data),
  \url{https://www.misoenergy.org/markets-and-operations/real-time--market-data/markets-displays/}
  (2025).

\bibitem{ReyesPremer2025_MinimalSensing_HPWH}
L.~D.~R. Premer, L.~Semmelmann, E.~A. Groll, D.~Ziviani, K.~J. Kircher, A
  minimal‑sensing predictive control strategy for heat pump water heaters:
  Field insights from cold climate operation, in: {2025 ASHRAE Annual
  Conference}, {ASHRAE}, 2025.

\bibitem{raspberrypi}
{{Raspberry Pi}}, {{Raspberry Pi} — cost‑effective, high‑performance
  computing for businesses and the home} (2025).

\bibitem{poweroutage2025}
B.~Woodburn, Electricity rates by state (may 2025) (2025).

\bibitem{NBI2023}
A.~Khanolkar, M.~Egolf, N.~Gabriel, Plug-in heat pump water heater field study
  findings and market commercialization recommendations, Tech. rep., {New
  Buildings Institute} (2023).

\bibitem{rewiringamerica2024}
{Rewiring America}, Upfront cost of home electrification (2024).

\bibitem{delaRosa_estimation}
L.~dela Rosa, A.~Gajjar, C.~Mande, M.~Ellis, Model-based hot water draw
  estimation in heat pump water heaters, {2024 ASHRAE Winter Conference} (01
  2024).

\end{thebibliography}

\end{document}